\documentclass{article}

 \usepackage[main, final]{neurips_2026}

\usepackage[utf8]{inputenc} 
\usepackage[T1]{fontenc}    
\usepackage{hyperref}       
\usepackage{url}            
\usepackage{booktabs}       
\usepackage{amsfonts}       
\usepackage{nicefrac}       
\usepackage{microtype}      
\usepackage{amsmath}
\usepackage{amssymb}

\usepackage{tikz}
\usepackage{adjustbox}
\usepackage{xcolor}
\usetikzlibrary{arrows.meta,positioning}
\usetikzlibrary{arrows.meta,positioning,fit}
\usetikzlibrary{calc,backgrounds}

\usepackage{booktabs}   
\usepackage{tabularx}   
\usepackage{array}      

\definecolor{ACEGraph}{HTML}{1F4E79}
\definecolor{ACEGraphFill}{HTML}{DCE8F5}
\definecolor{ACELoop}{HTML}{0F766E}
\definecolor{ACELoopFill}{HTML}{D9F2EE}
\definecolor{ACEControl}{HTML}{4B5563}
\definecolor{ACEControlFill}{HTML}{E9EDF2}
\definecolor{ACEInk}{HTML}{1F2933}
\definecolor{ACEVerify}{HTML}{2E7D32}
\definecolor{ACEVerifyFill}{HTML}{E4F3E6}

\title{Towards Agentic Cloud Engineering: Graph and Loop Engineering with a Zero-Trust Agent Harness}
\author{%
  Sagar Srinivas Sakhinana, Venkataramana Runkana \\
  \texttt{sagar.sakhinana@tcs.com, venkat.runkana@tcs.com} \\
  Tata Research Development and Design Centre
}

\begin{document}

\maketitle

\begin{abstract}
\vspace{-2mm}
Agentic AI is enabling cloud-based workflows in which autonomous agents reason over operational state, invoke authorized tools, modify software and infrastructure, deploy services, verify execution outcomes, and adapt across long-horizon, multi-step tasks. Engineering such workflows requires explicit mechanisms for workflow progression, constrained execution, failure recovery, and verifiable completion. We present \textbf{Agentic Cloud Workflow Engineering}, an agentic AI framework that transforms natural-language \emph{agentic cloud-engineering tasks} into validated code repositories and verified operational cloud deployments for automating cloud-based agentic workflows. The framework separates three complementary concerns: \emph{graph engineering} specifies long-horizon workflow progression and verification-dependent transitions; \emph{loop engineering} provides bounded diagnosis, repair or re-planning, retry, and re-verification; and \emph{agent harness engineering} enforces zero-trust execution through identity, authorization, policy-scoped capabilities, isolation, and runtime safeguards. Workflow progression and completion require machine-checkable repository, deployment, and runtime evidence, with recovery constrained by explicit operational bounds and termination criteria. We instantiate the framework on Google Cloud and evaluate repository completeness, controlled execution, evidence-gated progression, operational deployment, and bounded recovery. Experimental results show that executions terminate with either a verified operational cloud deployment or an auditable terminal failure under bounded recovery. The framework provides a unified engineering architecture for cloud-based workflows spanning Agentic DevOps, Agentic CloudOps, Agentic SRE/AIOps, Agentic SecOps, Agentic DataOps, Agentic MLOps/LLMOps, AgentOps, Agentic RAG/GraphRAG, and related cloud-engineering domains.
\vspace{-2mm}
\end{abstract}

\section{Introduction}
\label{sec:introduction}
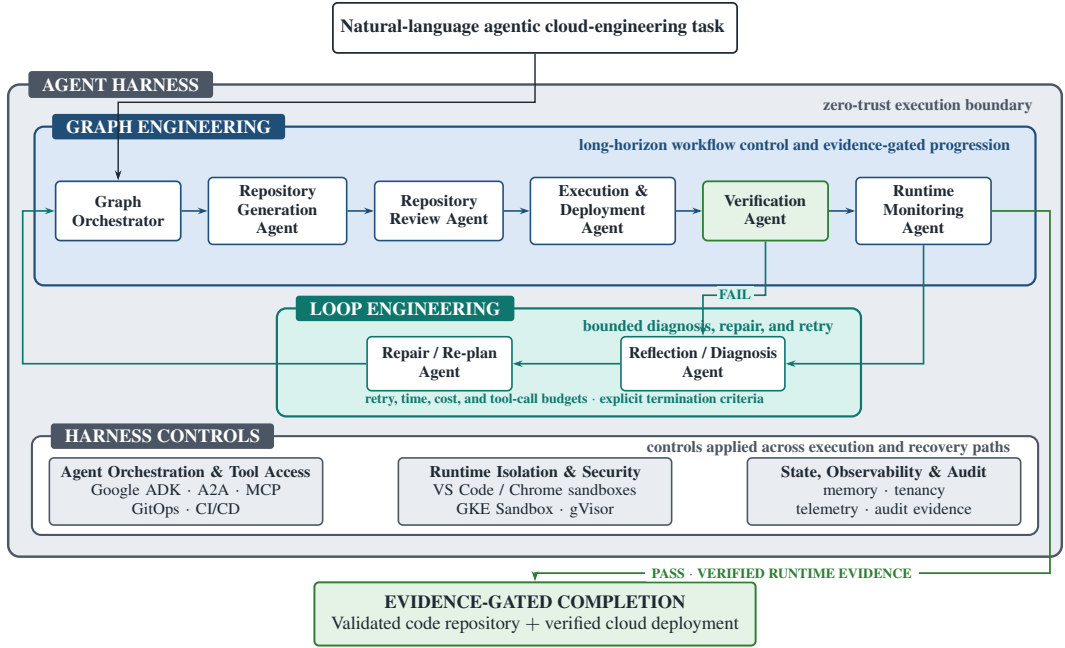
\begin{figure*}[ht!]
\centering

\begin{adjustbox}{max width=\textwidth}
\begin{tikzpicture}[x=1cm,y=1cm,font=\sffamily]


\node[
    draw=ACEInk,
    fill=white,
    rounded corners=2pt,
    line width=0.75pt,
    minimum width=4.7cm,
    minimum height=0.76cm,
    align=center,
    font=\footnotesize\bfseries,
    text=ACEInk
] (input) at (8,1.48)
{
    Natural-language agentic cloud-engineering task
};


\draw[
    draw=ACEControl,
    fill=ACEControlFill,
    rounded corners=5pt,
    line width=1.05pt
]
(0,0.62) rectangle (16,-6.55);

\node[
    anchor=west,
    fill=ACEControl,
    text=white,
    rounded corners=1.5pt,
    inner xsep=6pt,
    inner ysep=3pt,
    font=\footnotesize\bfseries
]
at (0.30,0.62)
{
    AGENT HARNESS
};

\node[
    anchor=east,
    text=ACEControl,
    font=\scriptsize\bfseries
]
at (15.68,0.29)
{
    zero-trust execution boundary
};


\draw[
    draw=ACEGraph,
    fill=ACEGraphFill,
    rounded corners=4pt,
    line width=0.95pt
]
(0.40,-0.02) rectangle (15.60,-2.43);

\node[
    anchor=west,
    fill=ACEGraph,
    text=white,
    rounded corners=1.5pt,
    inner xsep=6pt,
    inner ysep=2.7pt,
    font=\footnotesize\bfseries
]
at (0.66,-0.02)
{
    GRAPH ENGINEERING
};

\node[
    anchor=east,
    text=ACEGraph,
    font=\scriptsize\bfseries
]
at (15.34,-0.31)
{
    long-horizon workflow control and evidence-gated progression
};


\node[
    draw=ACEGraph,
    fill=white,
    rounded corners=2pt,
    line width=0.72pt,
    minimum width=1.90cm,
    text width=1.55cm,
    inner xsep=2pt,
    minimum height=0.90cm,
    align=center,
    font=\scriptsize\bfseries,
    text=ACEInk
] (orchestrator) at (1.67,-1.30)
{
    Graph\\
    Orchestrator
};


\node[
    draw=ACEGraph,
    fill=white,
    rounded corners=2pt,
    line width=0.68pt,
    minimum width=2.10cm,
    text width=1.75cm,
    inner xsep=2pt,
    minimum height=0.90cm,
    align=center,
    font=\scriptsize\bfseries,
    text=ACEInk
] (generate) at (4.09,-1.30)
{
    Repository\\
    Generation Agent
};


\node[
    draw=ACEGraph,
    fill=white,
    rounded corners=2pt,
    line width=0.68pt,
    minimum width=1.95cm,
    text width=1.60cm,
    inner xsep=2pt,
    minimum height=0.90cm,
    align=center,
    font=\scriptsize\bfseries,
    text=ACEInk
] (review) at (6.54,-1.30)
{
    Repository\\
    Review Agent
};


\node[
    draw=ACEGraph,
    fill=white,
    rounded corners=2pt,
    line width=0.68pt,
    minimum width=2.20cm,
    text width=1.90cm,
    inner xsep=2pt,
    minimum height=0.90cm,
    align=center,
    font=\scriptsize\bfseries,
    text=ACEInk
] (execute) at (9.03,-1.30)
{
    Execution \&\\
    Deployment Agent
};


\node[
    draw=ACEVerify,
    fill=ACEVerifyFill,
    rounded corners=2pt,
    line width=0.78pt,
    minimum width=1.90cm,
    text width=1.55cm,
    inner xsep=2pt,
    minimum height=0.90cm,
    align=center,
    font=\scriptsize\bfseries,
    text=ACEInk
] (verify) at (11.50,-1.30)
{
    Verification\\
    Agent
};


\node[
    draw=ACEGraph,
    fill=white,
    rounded corners=2pt,
    line width=0.68pt,
    minimum width=2.05cm,
    text width=1.70cm,
    inner xsep=2pt,
    minimum height=0.90cm,
    align=center,
    font=\scriptsize\bfseries,
    text=ACEInk
] (monitor) at (13.90,-1.30)
{
    Runtime\\
    Monitoring Agent
};


\draw[
    -{Stealth[length=1.55mm,width=1.0mm]},
    draw=ACEInk,
    line width=0.62pt
]
(input.south)
--
(8,0.35)
--
(1.67,0.35)
--
(orchestrator.north);


\draw[
    -{Stealth[length=1.5mm,width=1.0mm]},
    draw=ACEGraph,
    line width=0.68pt
]
(orchestrator.east) -- (generate.west);

\draw[
    -{Stealth[length=1.5mm,width=1.0mm]},
    draw=ACEGraph,
    line width=0.68pt
]
(generate.east) -- (review.west);

\draw[
    -{Stealth[length=1.5mm,width=1.0mm]},
    draw=ACEGraph,
    line width=0.68pt
]
(review.east) -- (execute.west);

\draw[
    -{Stealth[length=1.5mm,width=1.0mm]},
    draw=ACEGraph,
    line width=0.68pt
]
(execute.east) -- (verify.west);

\draw[
    -{Stealth[length=1.5mm,width=1.0mm]},
    draw=ACEGraph,
    line width=0.68pt
]
(verify.east) -- (monitor.west);


\draw[
    draw=ACELoop,
    fill=ACELoopFill,
    rounded corners=4pt,
    line width=0.95pt
]
(4.08,-4.42) rectangle (12.95,-2.78);

\node[
    anchor=west,
    fill=ACELoop,
    text=white,
    rounded corners=1.5pt,
    inner xsep=6pt,
    inner ysep=2.7pt,
    font=\footnotesize\bfseries
]
at (4.36,-2.78)
{
    LOOP ENGINEERING
};

\node[
    anchor=east,
    text=ACELoop,
    font=\scriptsize\bfseries
]
at (12.66,-3.07)
{
    bounded diagnosis, repair, and retry
};


\node[
    draw=ACELoop,
    fill=white,
    rounded corners=2pt,
    line width=0.70pt,
    minimum width=2.45cm,
    minimum height=0.82cm,
    align=center,
    font=\scriptsize\bfseries,
    text=ACEInk
] (reflect) at (10.55,-3.62)
{
    Reflection / Diagnosis\\
    Agent
};


\node[
    draw=ACELoop,
    fill=white,
    rounded corners=2pt,
    line width=0.70pt,
    minimum width=2.20cm,
    minimum height=0.82cm,
    align=center,
    font=\scriptsize\bfseries,
    text=ACEInk
] (repair) at (6.55,-3.62)
{
    Repair / Re-plan\\
    Agent
};


\draw[
    -{Stealth[length=1.5mm,width=1.0mm]},
    draw=ACELoop,
    line width=0.72pt
]
(verify.south)
--
(11.50,-2.58)
--
(10.55,-2.58)
--
(reflect.north);

\node[
    font=\tiny\bfseries,
    text=ACELoop,
    fill=ACELoopFill,
    inner sep=1pt
]
at (11.04,-2.56)
{
    FAIL
};


\draw[
    -{Stealth[length=1.5mm,width=1.0mm]},
    draw=ACELoop,
    line width=0.72pt
]
(monitor.south)
--
(13.90,-3.62)
--
(reflect.east);


\draw[
    -{Stealth[length=1.5mm,width=1.0mm]},
    draw=ACELoop,
    line width=0.72pt
]
(reflect.west) -- (repair.east);


\draw[
    -{Stealth[length=1.5mm,width=1.0mm]},
    draw=ACELoop,
    line width=0.72pt
]
(repair.west)
--
(0.22,-3.62)
--
(0.22,-1.30)
--
(orchestrator.west);

\node[
    font=\tiny\bfseries,
    text=ACELoop
]
at (8.45,-4.16)
{
    retry, time, cost, and tool-call budgets $\cdot$ explicit termination criteria
};


\draw[
    draw=ACEControl,
    fill=white,
    rounded corners=4pt,
    line width=0.90pt
]
(0.36,-6.22) rectangle (15.64,-4.72);

\node[
    anchor=west,
    fill=ACEControl,
    text=white,
    rounded corners=1.5pt,
    inner xsep=6pt,
    inner ysep=2.7pt,
    font=\footnotesize\bfseries
]
at (0.64,-4.72)
{
    HARNESS CONTROLS
};

\node[
    anchor=east,
    text=ACEControl,
    font=\scriptsize\bfseries
]
at (15.34,-4.90)
{
    controls applied across execution and recovery paths
};


\node[
    draw=ACEControl,
    fill=ACEControlFill,
    rounded corners=2pt,
    line width=0.68pt,
    minimum width=4.15cm,
    minimum height=0.92cm,
    text width=3.78cm,
    align=center,
    font=\scriptsize,
    text=ACEInk
] (tools) at (2.70,-5.56)
{
    \textbf{Agent Orchestration \& Tool Access}\\
    Google ADK $\cdot$ A2A $\cdot$ MCP\\
    GitOps $\cdot$ CI/CD
};


\node[
    draw=ACEControl,
    fill=ACEControlFill,
    rounded corners=2pt,
    line width=0.68pt,
    minimum width=4.15cm,
    minimum height=0.92cm,
    text width=3.78cm,
    align=center,
    font=\scriptsize,
    text=ACEInk
] (security) at (8.00,-5.56)
{
    \textbf{Runtime Isolation \& Security}\\
    VS Code / Chrome sandboxes\\
    GKE Sandbox $\cdot$ gVisor
};


\node[
    draw=ACEControl,
    fill=ACEControlFill,
    rounded corners=2pt,
    line width=0.68pt,
    minimum width=4.15cm,
    minimum height=0.92cm,
    text width=3.78cm,
    align=center,
    font=\scriptsize,
    text=ACEInk
] (state) at (13.30,-5.56)
{
    \textbf{State, Observability \& Audit}\\
    memory $\cdot$ tenancy\\
    telemetry $\cdot$ audit evidence
};


\node[
    draw=ACEVerify,
    fill=ACEVerifyFill,
    rounded corners=2pt,
    line width=0.85pt,
    minimum width=6.7cm,
    minimum height=0.96cm,
    align=center,
    font=\footnotesize,
    text=ACEInk
] (output) at (8,-7.42)
{
    \textbf{EVIDENCE-GATED COMPLETION}\\[-0.2mm]
    Validated code repository $+$ verified cloud deployment
};


\draw[
    -{Stealth[length=1.6mm,width=1.05mm]},
    draw=ACEVerify,
    line width=0.76pt
]
(monitor.east)
--
(15.82,-1.30)
--
(15.82,-6.80)
--
(8.00,-6.80)
--
(output.north);

\node[
    font=\tiny\bfseries,
    text=ACEVerify,
    fill=white,
    inner xsep=3pt,
    inner ysep=1.5pt,
    rounded corners=1pt
]
at (11.75,-6.80)
{
    PASS $\cdot$ VERIFIED RUNTIME EVIDENCE
};

\end{tikzpicture}
\end{adjustbox}
\caption{The figure shows an \textbf{Agentic Cloud Workflow Engineering} architecture for synthesizing, configuring, deploying, and verifying agentic workflows across DevOps, CloudOps, SRE/AIOps, SecOps, DataOps, MLOps/LLMOps, and related cloud-engineering domains. The Graph Orchestrator coordinates repository generation, review, execution and deployment, verification, and runtime monitoring through graph-structured workflows with bounded recovery. Verification failures or adverse runtime evidence trigger diagnosis, repair, re-planning, and retry, after which execution resumes through the Graph Orchestrator. The agent harness enforces per-action authorization, policy-scoped tool access, isolation, state, observability, audit, and evidence controls across execution and recovery paths. Repository generation, review, diagnosis, and repair use VS Code and Chrome sandboxes, whereas deployment and verification actions execute within the gVisor-isolated GKE Agent Sandbox. The software supply chain governs build, security validation, provenance, and controlled deployment. Evidence-gated completion requires a validated code repository, verified cloud deployment, and verified runtime evidence.}
\label{fig:ace-multiagent}
\vspace{-2mm}
\end{figure*}

\vspace{-2mm}
Organizations across industries—including Oil \& Gas, FMCG, Finance, Healthcare, Manufacturing, Telecommunications, Aerospace, Automotive, and Energy—increasingly rely on cloud platforms for elastic resource provisioning, fault tolerance, managed infrastructure, and scalable data and AI services. As cloud architectures and operational workloads increase in scale and complexity, agentic AI is increasingly applied to cloud-engineering workflows involving planning, implementation, deployment, operation, monitoring, verification, and adaptation~\citep{wang2024surveyagents,shetty2024autonomousclouds,chen2025aiopslab}. Agentic Cloud Engineering extends conventional cloud automation by incorporating goal-directed, tool-using agents into operational control loops. Unlike automation governed primarily by predefined execution paths, agents can interpret operational state, reason over objectives and constraints, select authorized tools and actions, evaluate execution feedback, and determine subsequent actions subject to explicit operational bounds. This operational pattern applies across multiple cloud-engineering domains. 
\textbf{(a) \emph{Agentic Software Engineering and DevOps}} covers repository analysis, code generation and modification, testing, debugging, repository repair, build execution, and software-delivery actions driven by execution feedback~\citep{yang2024sweagent,wang2025openhands}. \textbf{(b) \emph{Agentic Platform Engineering and CloudOps}} covers infrastructure-state analysis, infrastructure and configuration synthesis, provisioning, resource lifecycle management, and post-change validation~\citep{shetty2024autonomousclouds,yang2025cloudinfra}. \textbf{(c) \emph{Agentic AIOps and Site Reliability Engineering (SRE)}} covers telemetry analysis, incident triage, fault localization, root-cause analysis, mitigation planning, recovery actions, and post-recovery service validation~\citep{chen2025aiopslab,wang2024rcagent,chen2024rcacopilot}. \textbf{(d) \emph{Agentic Security Operations (SecOps)}} covers security-event analysis, threat detection, investigation, threat hunting, response planning, authorized containment, and post-response validation~\citep{lazer2026agenticcybersecurity}. \textbf{(e) \emph{Agentic Financial Operations (FinOps)}} covers cloud-cost and resource-utilization analysis, cost-anomaly detection, optimization planning, resource-rightsizing recommendations, and impact evaluation~\citep{vo2025finopsagent}. \textbf{(f) \emph{Agentic Data Operations (DataOps)}} covers data-pipeline synthesis and modification, data transformation, execution diagnosis, recovery and replay, and output validation~\citep{siva2026kraig}. \textbf{(g) \emph{Agentic MLOps and LLMOps}} covers agent-directed model and pipeline experimentation, evaluation, pipeline modification, operational monitoring, failure diagnosis, retraining, model selection, deployment, promotion, and rollback based on observed lifecycle state~\citep{nam2025mlestar,hu2025mlopsmonitoring}. \textbf{(h) \emph{Agent Operations (AgentOps)}} covers tracing, evaluation, monitoring, logging, execution analysis, and lifecycle management of agents and multi-agent workflows~\citep{dong2024agentops}. \textbf{(i) \emph{Agentic RAG and GraphRAG}} covers agent-directed query formulation, retrieval planning, source and tool selection, iterative evidence acquisition, graph-based retrieval, synthesis, and grounding verification~\citep{lewis2020rag,edge2024graphrag}. \textbf{(j) \emph{Agentic Database Operations}} covers database-state and workload analysis, anomaly and performance diagnosis, root-cause analysis, corrective-action planning, authorized interventions, and post-action validation~\citep{zhou2024dbot}. \textbf{(k) \emph{Agentic Network Operations (NetOps)}} covers network-telemetry analysis, fault localization, configuration synthesis, change planning, authorized configuration actions, bounded recovery, and post-change validation~\citep{bilal2026agenticnetops}. \textbf{(l) \emph{Agentic Identity and Access Management (IAM)}} covers identity-context analysis, delegated authorization, dynamic permission scoping, credential selection, access-control decisions, and least-privilege enforcement for agent actions~\citep{south2025delegation}. \textbf{(m) \emph{Agentic Governance and Compliance}} covers policy interpretation, policy-as-code generation and validation, compliance assessment, policy-constrained action planning, approval routing, and corrective configuration generation~\citep{romeo2025arpaccino}. \textbf{(n) \emph{Agentic Inference Operations}} covers inference-state analysis, request scheduling and placement, dynamic batching, cache management, load distribution, and accelerator-resource allocation for agent workloads~\citep{guo2026saga}. Across these domains, diverse cloud-engineering objectives must be realized through closed-loop agentic workflows that iteratively Observe $\rightarrow$ Reason $\rightarrow$ Plan $\rightarrow$ Act $\rightarrow$ Verify $\rightarrow$ Adapt based on system state and execution evidence~\citep{yao2023react,shinn2023reflexion,wang2024surveyagents}. \textbf{Agentic Cloud Workflow Engineering} addresses the synthesis, configuration, deployment, and verification of such workflows. For example, a natural-language Agentic Cloud Engineering task such as ``Build and deploy a multi-tenant Agentic RAG platform for retrieving, reasoning over, synthesizing, and verifying enterprise knowledge'' requires a deployable implementation comprising retrieval, reasoning and synthesis, and verification agents, together with tenant isolation, Role-Based Access Control (RBAC), Attribute-Based Access Control (ABAC), document-level authorization, Personally Identifiable Information (PII) protection, guardrails, provenance, and grounded-response verification. The framework must synthesize and validate the required repository artifacts, deploy the resulting services, verify end-to-end runtime behavior, and recover from failed verification across long-horizon, multi-step workflows~\citep{yao2023react,shinn2023reflexion,shetty2024autonomousclouds,chen2025aiopslab}. Realizing such requirements therefore demands coordinated control over \emph{where workflow execution proceeds}, \emph{how execution failures are diagnosed and corrected}, and \emph{what permissions, tools, isolation boundaries, and policies constrain agent actions}; these requirements motivate graph engineering, bounded loop engineering, and agent harness engineering, respectively.
We formulate \textbf{Agentic Cloud Workflow Engineering} through three complementary abstractions---\emph{graph engineering}, \emph{loop engineering}, and \emph{agent harness engineering}---as illustrated in Fig.~\ref{fig:ace-multiagent}. \emph{Graph engineering} specifies workflow states, transitions, branches, dependencies, approvals, recovery paths, and terminal conditions for long-horizon execution and evidence-gated progression~\citep{openai2025practicalagents,anthropic2024effectiveagents}. In Fig.~\ref{fig:ace-multiagent}, the Graph Orchestrator coordinates a natural-language Agentic Cloud Engineering task across repository generation, review, execution and deployment, verification, and runtime monitoring. \emph{Loop engineering} governs bounded diagnosis, repair or re-planning, retry, and re-verification following failed verification or adverse runtime evidence, subject to limits on attempts, time, cost, and tool calls~\citep{anthropic2024effectiveagents,anthropic2025effectiveharnesses}. Following repair or re-planning, the Graph Orchestrator evaluates the updated workflow state and selects the applicable graph transition. \emph{Agent harness engineering} defines the zero-trust execution boundary for tool invocation and cloud-engineering actions under identity, least-privilege authorization, policy, isolation, state, observability, audit, and evidence controls~\citep{openai2026agentssdk,openai2026sandboxagents,anthropic2025effectiveharnesses}. Together, these abstractions transform natural-language Agentic Cloud Engineering tasks into validated code repositories and verified cloud deployments, with completion gated by repository, deployment, and runtime evidence.

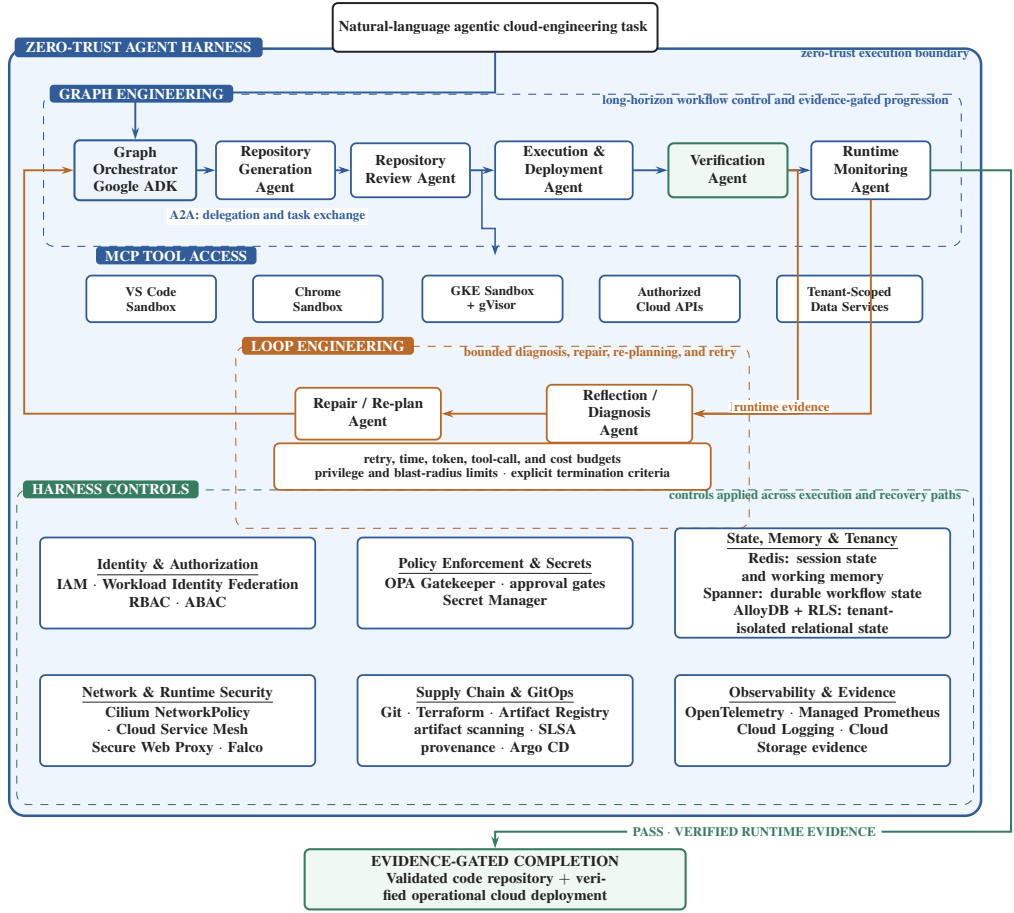
\begin{figure*}[ht!]
\centering

\definecolor{ACEBlue}{RGB}{48,93,154}
\definecolor{ACEBlueFill}{RGB}{239,245,252}
\definecolor{ACEGreen}{RGB}{54,125,96}
\definecolor{ACEGreenFill}{RGB}{239,248,243}
\definecolor{ACEOrange}{RGB}{184,104,40}
\definecolor{ACEInk}{RGB}{35,35,35}

\begin{tikzpicture}[
    scale=0.70,
    transform shape,
    request/.style={
        draw=ACEInk,
        fill=white,
        rounded corners=2pt,
        line width=0.75pt,
        align=center,
        text=ACEInk,
        minimum height=0.78cm
    },
    process/.style={
        draw=ACEBlue,
        fill=white,
        rounded corners=2pt,
        line width=0.68pt,
        align=center,
        text=ACEInk
    },
    graphcontrol/.style={
        draw=ACEBlue,
        fill=ACEBlueFill,
        rounded corners=2pt,
        line width=0.78pt,
        align=center,
        text=ACEInk
    },
    verification/.style={
        draw=ACEGreen,
        fill=ACEGreenFill,
        rounded corners=2pt,
        line width=0.78pt,
        align=center,
        text=ACEInk
    },
    mainarrow/.style={
        -{Stealth[length=1.6mm,width=1.0mm]},
        draw=ACEBlue,
        line width=0.72pt
    },
    toolarrow/.style={
        -{Stealth[length=1.45mm,width=0.95mm]},
        draw=ACEBlue,
        line width=0.58pt
    },
    looparrow/.style={
        -{Stealth[length=1.6mm,width=1.0mm]},
        draw=ACEOrange,
        line width=0.72pt
    },
    looplabel/.style={
        font=\scriptsize\bfseries,
        text=ACEOrange,
        fill=white,
        inner sep=1pt
    },
    agent/.style={
        process,
        minimum width=2.25cm,
        text width=1.95cm,
        minimum height=1.02cm,
        font=\footnotesize\bfseries,
        align=center,
        inner sep=2.5pt
    },
    mcpbox/.style={
        process,
        minimum width=2.45cm,
        text width=2.10cm,
        minimum height=0.88cm,
        font=\scriptsize\bfseries,
        align=center,
        inner sep=2pt
    },
    controlbox/.style={
        process,
        minimum width=5.20cm,
        text width=4.82cm,
        minimum height=1.72cm,
        font=\footnotesize\bfseries,
        align=center,
        inner sep=2.5pt
    },
    loopagent/.style={
        process,
        draw=ACEOrange,
        minimum width=2.75cm,
        text width=2.40cm,
        minimum height=0.98cm,
        font=\footnotesize\bfseries,
        align=center,
        inner sep=2.5pt
    },
    smallnote/.style={
        font=\scriptsize\bfseries,
        inner sep=1.3pt
    }
]


\node[
    request,
    minimum width=5.8cm,
    minimum height=0.92cm,
    font=\footnotesize\bfseries
] (req) at (9.10,1.95) {
    Natural-language agentic cloud-engineering task
};


\node[
    graphcontrol,
    minimum width=2.30cm,
    text width=2.00cm,
    minimum height=1.06cm,
    font=\footnotesize\bfseries,
    align=center
] (orch) at (2.30,-0.75) {
    \textbf{Graph\\Orchestrator}\\[-0.2mm]
    Google ADK
};

\node[agent] (generate) at (4.95,-0.75) {
    \textbf{Repository\\Generation Agent}
};

\node[agent] (review) at (7.50,-0.75) {
    \textbf{Repository\\Review Agent}
};

\node[
    agent,
    minimum width=2.60cm,
    text width=2.25cm
] (execute) at (10.40,-0.75) {
    \textbf{Execution \&\\Deployment Agent}
};

\node[
    verification,
    minimum width=2.25cm,
    text width=1.95cm,
    minimum height=1.02cm,
    font=\footnotesize\bfseries,
    align=center
] (verify) at (13.50,-0.75) {
    \textbf{Verification\\Agent}
};

\node[agent] (monitor) at (16.20,-0.75) {
    \textbf{Runtime\\Monitoring Agent}
};


\draw[mainarrow]
    (req.south)
    --
    (9.10,0.72)
    --
    (2.30,0.72)
    --
    (orch.north);


\draw[mainarrow] (orch.east) -- (generate.west);
\draw[mainarrow] (generate.east) -- (review.west);
\draw[mainarrow] (review.east) -- (execute.west);
\draw[mainarrow] (execute.east) -- (verify.west);
\draw[mainarrow] (verify.east) -- (monitor.west);


\node[
    smallnote,
    text=ACEBlue,
    fill=white
] at (4.80,-1.62) {
    A2A: delegation and task exchange
};


\node[
    anchor=west,
    fill=ACEBlue,
    text=white,
    rounded corners=1.5pt,
    inner xsep=5pt,
    inner ysep=2pt,
    font=\footnotesize\bfseries
] (mcplabel) at (1.55,-2.36) {
    MCP TOOL ACCESS
};


\node[
    mcpbox
] (vscode) at (2.60,-3.18) {
    VS Code\\
    Sandbox
};

\node[
    mcpbox
] (chrome) at (5.75,-3.18) {
    Chrome\\
    Sandbox
};

\node[
    mcpbox,
    minimum width=2.65cm,
    text width=2.30cm
] (gke) at (9.05,-3.18) {
    GKE Sandbox\\
    + gVisor
};

\node[
    mcpbox,
    minimum width=2.65cm,
    text width=2.30cm
] (cloudapi) at (12.40,-3.18) {
    Authorized\\
    Cloud APIs
};

\node[
    mcpbox,
    minimum width=2.75cm,
    text width=2.40cm
] (datasvc) at (15.80,-3.18) {
    Tenant-Scoped\\
    Data Services
};


\begin{scope}[on background layer]

\node[
    draw=ACEBlue,
    dashed,
    rounded corners=4pt,
    inner xsep=3.5mm,
    inner ysep=4mm,
    fit={
        (vscode)
        (chrome)
        (gke)
        (cloudapi)
        (datasvc)
    }
] (mcpboundary) {};

\end{scope}


\coordinate (graphaccess) at (8.85,-0.75);

\draw[toolarrow]
    (graphaccess)
    --
    (8.85,-1.85)
    --
    (9.10,-1.85)
    --
    (9.10,-2.36);


\node[
    loopagent
] (reflect) at (11.45,-5.35) {
    \textbf{Reflection / Diagnosis\\Agent}
};

\node[
    loopagent
] (repair) at (6.70,-5.35) {
    \textbf{Repair / Re-plan\\Agent}
};


\node[
    process,
    draw=ACEOrange,
    minimum width=8.25cm,
    text width=7.85cm,
    minimum height=0.88cm,
    font=\scriptsize\bfseries,
    align=center
] (bounds) at (9.05,-6.35) {
    retry, time, token, tool-call, and cost budgets\\[-0.1mm]
    privilege and blast-radius limits $\cdot$ explicit termination criteria
};


\draw[looparrow]
    (verify.east)
    --
    (14.82,-0.75)
    --
    (14.82,-5.35)
    --
    node[
        looplabel,
        above
    ] {FAIL}
    (reflect.east);


\draw[looparrow]
    (monitor.south)
    --
    (16.20,-5.35)
    --
    node[
        looplabel,
        above
    ] {runtime evidence}
    (reflect.east);


\draw[looparrow]
    (reflect.west)
    --
    (repair.east);

%

\coordinate (recoveryleft) at (0.20,-5.35);
\coordinate (recoveryup)   at (0.20,-0.75);

\draw[looparrow]
    (repair.west)
    --
    (recoveryleft)
    --
    (recoveryup)
    --
    (orch.west);


\node[
    controlbox
] (identity) at (3.10,-8.55) {
    \underline{\textbf{Identity \& Authorization}}\\[0.45mm]
    IAM $\cdot$ Workload Identity Federation\\[-0.1mm]
    RBAC $\cdot$ ABAC
};

\node[
    controlbox
] (policy) at (9.10,-8.55) {
    \underline{\textbf{Policy Enforcement \& Secrets}}\\[0.45mm]
    OPA Gatekeeper $\cdot$ approval gates\\[-0.1mm]
    Secret Manager
};

\node[
    controlbox
] (state) at (15.10,-8.55) {
    \underline{\textbf{State, Memory \& Tenancy}}\\[0.45mm]
    Redis: session state and working memory\\[-0.1mm]
    Spanner: durable workflow state\\[-0.1mm]
    AlloyDB + RLS: tenant-isolated relational state
};


\node[
    controlbox
] (network) at (3.10,-11.15) {
    \underline{\textbf{Network \& Runtime Security}}\\[0.45mm]
    Cilium NetworkPolicy $\cdot$ Cloud Service Mesh\\[-0.1mm]
    Secure Web Proxy $\cdot$ Falco
};

\node[
    controlbox
] (supplychain) at (9.10,-11.15) {
    \underline{\textbf{Supply Chain \& GitOps}}\\[0.45mm]
    Git $\cdot$ Terraform $\cdot$ Artifact Registry\\[-0.1mm]
    artifact scanning $\cdot$ SLSA provenance $\cdot$ Argo CD
};

\node[
    controlbox
] (observe) at (15.10,-11.15) {
    \underline{\textbf{Observability \& Evidence}}\\[0.45mm]
    OpenTelemetry $\cdot$ Managed Prometheus\\[-0.1mm]
    Cloud Logging $\cdot$ Cloud Storage evidence
};


\node[
    verification,
    minimum width=7.20cm,
    text width=6.75cm,
    minimum height=1.08cm,
    font=\footnotesize\bfseries,
    align=center
] (complete) at (9.10,-14.15) {
    \textbf{EVIDENCE-GATED COMPLETION}\\[-0.15mm]
    Validated code repository $+$
    verified operational cloud deployment
};


\draw[
    -{Stealth[length=1.6mm,width=1.0mm]},
    draw=ACEGreen,
    line width=0.76pt
]
    (monitor.east)
    --
    (18.85,-0.75)
    --
    (18.85,-13.25)
    --
    (9.10,-13.25)
    --
    (complete.north);

\node[
    smallnote,
    text=ACEGreen,
    fill=white,
    inner xsep=3pt,
    inner ysep=1.5pt,
    rounded corners=1pt
] at (14.00,-13.25) {
    PASS $\cdot$ VERIFIED RUNTIME EVIDENCE
};


\coordinate (harnesstop) at (9.10,1.12);
\coordinate (harnessbottom) at (9.10,-12.52);


\begin{scope}[on background layer]


\node[
    draw=ACEBlue,
    fill=ACEBlueFill,
    rounded corners=6pt,
    line width=0.95pt,
    inner xsep=4mm,
    inner ysep=3mm,
    fit={
        (harnesstop)
        (harnessbottom)
        (orch)
        (generate)
        (review)
        (execute)
        (verify)
        (monitor)
        (mcplabel)
        (mcpboundary)
        (vscode)
        (chrome)
        (gke)
        (cloudapi)
        (datasvc)
        (reflect)
        (repair)
        (bounds)
        (identity)
        (policy)
        (state)
        (network)
        (supplychain)
        (observe)
    }
] (harnessboundary) {};


\node[
    draw=ACEBlue,
    dashed,
    rounded corners=4pt,
    inner xsep=4mm,
    inner ysep=6mm,
    fit={
        (orch)
        (generate)
        (review)
        (execute)
        (verify)
        (monitor)
    }
] (graphboundary) {};


\node[
    draw=ACEOrange,
    dashed,
    rounded corners=4pt,
    inner xsep=5mm,
    inner ysep=5mm,
    fit={
        (reflect)
        (repair)
        (bounds)
    }
] (loopboundary) {};


\node[
    draw=ACEGreen,
    dashed,
    rounded corners=4pt,
    inner xsep=3mm,
    inner ysep=5mm,
    fit={
        (identity)
        (policy)
        (state)
        (network)
        (supplychain)
        (observe)
    }
] (controlboundary) {};

\end{scope}


\node[
    anchor=west,
    fill=ACEBlue,
    text=white,
    rounded corners=1.5pt,
    inner xsep=6pt,
    inner ysep=2.5pt,
    font=\footnotesize\bfseries
] at ($(harnessboundary.north west)+(0.12cm,0)$) {
    ZERO-TRUST AGENT HARNESS
};

\node[
    anchor=east,
    font=\scriptsize\bfseries,
    text=ACEBlue
] at ($(harnessboundary.north east)+(-0.12cm,-0.13cm)$) {
    zero-trust execution boundary
};

\node[
    anchor=west,
    fill=ACEBlue,
    text=white,
    rounded corners=1.5pt,
    inner xsep=5pt,
    inner ysep=2pt,
    font=\footnotesize\bfseries
] at ($(graphboundary.north west)+(0.12cm,0)$) {
    GRAPH ENGINEERING
};

\node[
    anchor=east,
    font=\scriptsize\bfseries,
    text=ACEBlue
] at ($(graphboundary.north east)+(-0.12cm,-0.13cm)$) {
    long-horizon workflow control and evidence-gated progression
};

\node[
    anchor=west,
    fill=ACEOrange,
    text=white,
    rounded corners=1.5pt,
    inner xsep=5pt,
    inner ysep=2pt,
    font=\footnotesize\bfseries
] at ($(loopboundary.north west)+(0.12cm,0)$) {
    LOOP ENGINEERING
};

\node[
    anchor=east,
    font=\scriptsize\bfseries,
    text=ACEOrange
] at ($(loopboundary.north east)+(-0.12cm,-0.13cm)$) {
    bounded diagnosis, repair, re-planning, and retry
};

\node[
    anchor=west,
    fill=ACEGreen,
    text=white,
    rounded corners=1.5pt,
    inner xsep=5pt,
    inner ysep=2pt,
    font=\footnotesize\bfseries
] at ($(controlboundary.north west)+(0.12cm,0)$) {
    HARNESS CONTROLS
};

\node[
    anchor=east,
    font=\scriptsize\bfseries,
    text=ACEGreen
] at ($(controlboundary.north east)+(-0.12cm,-0.13cm)$) {
    controls applied across execution and recovery paths
};

\end{tikzpicture}
\caption{The figure shows the \textbf{Agentic Cloud Workflow Engineering} architecture integrating graph engineering, loop engineering, and a zero-trust agent harness. The Graph Orchestrator uses Google Agent Development Kit (ADK) to coordinate long-horizon workflow execution, while Agent2Agent (A2A) supports inter-agent delegation and task exchange and Model Context Protocol (MCP) provides scoped access to sandboxes, authorized cloud APIs, and tenant-scoped data services. Verification failures or adverse runtime evidence activate bounded diagnosis, repair, re-planning, retry, and re-verification. The agent harness enforces identity and authorization, policy, isolation, state and tenancy, network and runtime security, software supply-chain, observability, and evidence controls across execution and recovery paths. Evidence-gated completion requires a validated code repository, verified operational cloud deployment, and verified runtime evidence.}
\label{fig:agentic-cloud-framework}
\vspace{-6mm}
\end{figure*}

Building on the framework-level abstraction in Fig.~\ref{fig:ace-multiagent}, Fig.~\ref{fig:agentic-cloud-framework} presents a system-level realization of \textbf{Agentic Cloud Workflow Engineering} on Google Cloud. The graph, loop, and agent harness abstractions are cloud-platform independent, while the architecture instantiates them using Google Agent Development Kit (ADK) for workflow orchestration, Agent2Agent (A2A) for inter-agent delegation and task exchange, Model Context Protocol (MCP) for scoped tool access, isolated execution environments, and zero-trust controls spanning identity, authorization, policy enforcement, state, runtime security, observability, and evidence capture. Verification failures or adverse runtime evidence trigger bounded recovery comprising diagnosis, repair or re-planning, retry, and re-verification. Evidence-gated completion requires a validated code repository, a verified operational cloud deployment, and verified runtime evidence. Our contributions are as follows:

\vspace{-1mm}
\begin{itemize}
\item We formulate \emph{Agentic Cloud Workflow Engineering} through three complementary abstractions: \emph{graph engineering} for long-horizon workflow control and evidence-gated progression, \emph{loop engineering} for bounded diagnosis and recovery, and \emph{agent harness engineering} for policy-constrained agent execution.
\vspace{-1mm}
\item We design a \emph{zero-trust agent harness} that constrains agent actions through scoped tool access, workload identity, least-privilege authorization, policy enforcement, tenant isolation, sandboxed execution, runtime security, state management, observability, audit, and evidence capture.
\vspace{-1mm}
\item We define \emph{bounded recovery}, in which verification failures and adverse runtime evidence trigger diagnosis, repair or re-planning, retry, and re-verification under explicit retry, time, token, tool-call, cost, privilege, and blast-radius limits with defined termination criteria.
\vspace{-1mm}
\item We define \emph{evidence-gated progression and completion}, in which workflow transitions require machine-checkable verification evidence and successful termination requires validated repository, deployment, and runtime evidence rather than agent-reported task completion.
\end{itemize}

The subsequent sections detail the problem formulation, framework, experimental setup, benchmark suite, and empirical evaluation.
\vspace{-2mm}

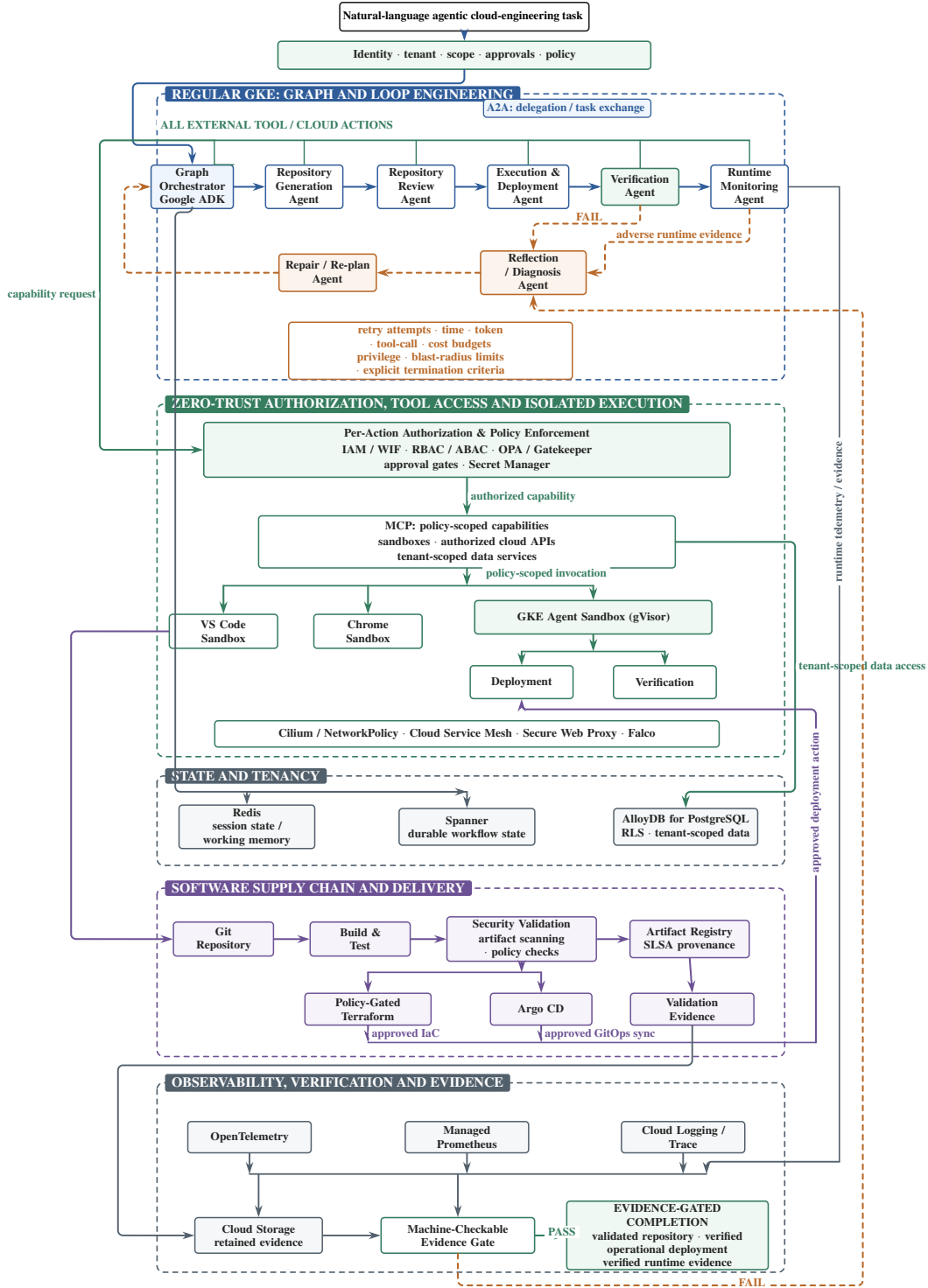
\begin{figure*}[ht!]
\centering

\definecolor{ACEBlue}{RGB}{48,93,154}
\definecolor{ACEBlueFill}{RGB}{239,245,252}
\definecolor{ACEGreen}{RGB}{54,125,96}
\definecolor{ACEGreenFill}{RGB}{239,248,243}
\definecolor{ACEOrange}{RGB}{184,104,40}
\definecolor{ACEOrangeFill}{RGB}{252,244,236}
\definecolor{ACEPurple}{RGB}{108,82,148}
\definecolor{ACEPurpleFill}{RGB}{247,243,251}
\definecolor{ACESlate}{RGB}{82,96,108}
\definecolor{ACESlateFill}{RGB}{245,247,248}
\definecolor{ACEInk}{RGB}{35,35,35}

\begin{adjustbox}{max width=\textwidth}

\begin{tikzpicture}[
    scale=0.59,
    x=0.95cm,
    y=0.64cm,
    transform shape,
    line cap=round,
    line join=round,
    every node/.append style={
        execute at begin node={
            \hyphenpenalty=10000
            \exhyphenpenalty=10000
            \relax
        }
    },
    box/.style={
        draw=ACEBlue,
        fill=white,
        rounded corners=2pt,
        line width=0.80pt,
        align=center,
        text=ACEInk,
        font=\large\bfseries,
        inner sep=3.2pt
    },
    ctl/.style={
        box,
        fill=ACEBlueFill
    },
    secure/.style={
        draw=ACEGreen,
        fill=ACEGreenFill,
        rounded corners=2pt,
        line width=0.84pt,
        align=center,
        text=ACEInk,
        font=\large\bfseries,
        inner sep=3.2pt
    },
    loopbox/.style={
        draw=ACEOrange,
        fill=ACEOrangeFill,
        rounded corners=2pt,
        line width=0.84pt,
        align=center,
        text=ACEInk,
        font=\large\bfseries,
        inner sep=3.2pt
    },
    capability/.style={
        draw=ACEGreen,
        fill=white,
        rounded corners=2pt,
        line width=0.82pt,
        align=center,
        text=ACEInk,
        font=\large\bfseries,
        inner sep=3.2pt
    },
    store/.style={
        draw=ACESlate,
        fill=ACESlateFill,
        rounded corners=2pt,
        line width=0.80pt,
        align=center,
        text=ACEInk,
        font=\large\bfseries,
        inner sep=3.2pt
    },
    pipebox/.style={
        draw=ACEPurple,
        fill=ACEPurpleFill,
        rounded corners=2pt,
        line width=0.80pt,
        align=center,
        text=ACEInk,
        font=\large\bfseries,
        inner sep=3.2pt
    },
    obsbox/.style={
        draw=ACESlate,
        fill=ACESlateFill,
        rounded corners=2pt,
        line width=0.80pt,
        align=center,
        text=ACEInk,
        font=\large\bfseries,
        inner sep=3.2pt
    },
    verifybox/.style={
        draw=ACEGreen,
        fill=white,
        rounded corners=2pt,
        line width=0.84pt,
        align=center,
        text=ACEInk,
        font=\large\bfseries,
        inner sep=3.2pt
    },
    title/.style={
        font=\Large\bfseries,
        text=white,
        inner xsep=7pt,
        inner ysep=3pt,
        rounded corners=1.5pt
    },
    arrow/.style={
        -{Stealth[length=2.70mm,width=1.72mm]},
        draw=ACEBlue,
        line width=1.08pt,
        rounded corners=3pt
    },
    securearrow/.style={
        -{Stealth[length=2.70mm,width=1.72mm]},
        draw=ACEGreen,
        line width=1.10pt,
        rounded corners=3pt
    },
    statearrow/.style={
        -{Stealth[length=2.60mm,width=1.62mm]},
        draw=ACESlate,
        line width=1.00pt,
        rounded corners=3pt
    },
    deliveryarrow/.style={
        -{Stealth[length=2.70mm,width=1.72mm]},
        draw=ACEPurple,
        line width=1.08pt,
        rounded corners=3pt
    },
    feedback/.style={
        -{Stealth[length=2.70mm,width=1.72mm]},
        draw=ACEOrange,
        line width=1.10pt,
        dashed,
        rounded corners=3pt
    },
    note/.style={
        font=\large\bfseries,
        text=ACEInk,
        fill=white,
        inner xsep=3.5pt,
        inner ysep=1.8pt
    }
]


\node[
    draw=ACEInk,
    fill=white,
    rounded corners=2pt,
    line width=0.86pt,
    minimum width=7.40cm,
    minimum height=1.04cm,
    font=\large\bfseries,
    align=center
] (req) at (12.45,4.60) {
    Natural-language agentic cloud-engineering task
};

\node[
    secure,
    minimum width=13.60cm,
    text width=12.90cm,
    minimum height=1.02cm
] (context) at (12.45,2.40) {
    Identity $\cdot$ tenant $\cdot$ scope $\cdot$ approvals $\cdot$ policy
};

\draw[arrow]
    (req.south) -- (context.north);


\draw[
    draw=ACEBlue,
    dashed,
    rounded corners=4pt,
    line width=0.88pt
]
    (0.65,0.20) rectangle (24.75,-16.10);

\draw[
    draw=ACEGreen,
    dashed,
    rounded corners=4pt,
    line width=0.88pt
]
    (0.65,-17.50) rectangle (24.75,-37.60);

\draw[
    draw=ACESlate,
    dashed,
    rounded corners=4pt,
    line width=0.88pt
]
    (0.65,-38.70) rectangle (24.75,-43.80);

\draw[
    draw=ACEPurple,
    dashed,
    rounded corners=4pt,
    line width=0.88pt
]
    (0.65,-45.10) rectangle (24.75,-54.70);

\draw[
    draw=ACESlate,
    dashed,
    rounded corners=4pt,
    line width=0.88pt
]
    (0.65,-56.20) rectangle (24.75,-67.00);


\node[
    title,
    fill=ACEBlue,
    anchor=west
] at (0.95,0.20) {
    REGULAR GKE: GRAPH AND LOOP ENGINEERING
};

\node[
    title,
    fill=ACEGreen,
    anchor=west
] at (0.95,-17.50) {
    ZERO-TRUST AUTHORIZATION, TOOL ACCESS AND ISOLATED EXECUTION
};

\node[
    title,
    fill=ACESlate,
    anchor=west
] at (0.95,-38.70) {
    STATE AND TENANCY
};

\node[
    title,
    fill=ACEPurple,
    anchor=west
] at (0.95,-45.10) {
    SOFTWARE SUPPLY CHAIN AND DELIVERY
};

\node[
    title,
    fill=ACESlate,
    anchor=west
] at (0.95,-56.20) {
    OBSERVABILITY, VERIFICATION AND EVIDENCE
};


\node[
    ctl,
    minimum width=3.00cm,
    text width=2.64cm,
    minimum height=1.46cm
] (orch) at (2.00,-5.10) {
    \textbf{Graph}\\
    \textbf{Orchestrator}\\
    Google ADK
};

\node[
    box,
    minimum width=2.82cm,
    text width=2.42cm,
    minimum height=1.46cm
] (generate) at (6.30,-5.10) {
    \textbf{Repository\\Generation\\Agent}
};

\node[
    box,
    minimum width=2.82cm,
    text width=2.42cm,
    minimum height=1.46cm
] (review) at (10.60,-5.10) {
    \textbf{Repository\\Review\\Agent}
};

\node[
    box,
    minimum width=2.96cm,
    text width=2.54cm,
    minimum height=1.46cm
] (execute) at (14.90,-5.10) {
    \textbf{Execution \&\\Deployment\\Agent}
};

\node[
    secure,
    minimum width=2.82cm,
    text width=2.42cm,
    minimum height=1.34cm
] (verify) at (19.20,-5.10) {
    \textbf{Verification\\Agent}
};

\node[
    box,
    minimum width=2.82cm,
    text width=2.42cm,
    minimum height=1.46cm
] (monitor) at (23.40,-5.10) {
    \textbf{Runtime\\Monitoring\\Agent}
};


\draw[arrow]
    (context.south)
    -- (12.45,0.80)
    -- (-0.28,0.80)
    -- (-0.28,-2.72)
    -- (2.00,-2.72)
    -- (orch.north);


\draw[arrow]
    (orch.east) -- (generate.west);

\draw[arrow]
    (generate.east) -- (review.west);

\draw[arrow]
    (review.east) -- (execute.west);

\draw[arrow]
    (execute.east) -- (verify.west);

\draw[arrow]
    (verify.east) -- (monitor.west);


\node[
    draw=ACEBlue,
    fill=ACEBlueFill,
    rounded corners=2pt,
    line width=0.70pt,
    minimum width=5.60cm,
    minimum height=0.66cm,
    font=\large\bfseries,
    align=center,
    text=ACEBlue
] (a2a) at (16.40,-0.62) {
    A2A: delegation / task exchange
};


\draw[
    draw=ACEGreen,
    line width=0.94pt
]
    (2.85,-2.45) -- (23.40,-2.45);

\draw[
    draw=ACEGreen,
    line width=0.76pt
]
    (2.85,-2.45) |- (orch.north east);

\draw[
    draw=ACEGreen,
    line width=0.76pt
]
    (generate.north) -- (6.30,-2.45);

\draw[
    draw=ACEGreen,
    line width=0.76pt
]
    (review.north) -- (10.60,-2.45);

\draw[
    draw=ACEGreen,
    line width=0.76pt
]
    (execute.north) -- (14.90,-2.45);

\draw[
    draw=ACEGreen,
    line width=0.76pt
]
    (verify.north) -- (19.20,-2.45);

\draw[
    draw=ACEGreen,
    line width=0.76pt
]
    (monitor.north) -- (23.40,-2.45);

\node[
    note,
    text=ACEGreen
] at (5.25,-1.55) {
    ALL EXTERNAL TOOL / CLOUD ACTIONS
};



\node[
    loopbox,
    minimum width=3.55cm,
    text width=3.08cm,
    minimum height=1.36cm
] (repair) at (7.20,-10.00) {
    \textbf{Repair / Re-plan\\Agent}
};

\node[
    loopbox,
    minimum width=3.85cm,
    text width=3.38cm,
    minimum height=1.36cm
] (reflect) at (15.10,-10.00) {
    \textbf{Reflection / Diagnosis\\Agent}
};


\node[
    draw=ACEOrange,
    fill=white,
    rounded corners=2pt,
    line width=0.80pt,
    minimum width=10.35cm,
    text width=9.70cm,
    minimum height=1.10cm,
    font=\large\bfseries,
    align=center,
    text=ACEOrange,
    inner sep=3.4pt
] (bounds) at (11.15,-14.45) {
    retry attempts $\cdot$ time $\cdot$ token $\cdot$
    tool-call $\cdot$ cost budgets\\[-0.1mm]
    privilege $\cdot$ blast-radius limits $\cdot$
    explicit termination criteria
};


\draw[feedback]
    (verify.south)
    -- (19.20,-7.20)
    -- (15.10,-7.20)
    -- (reflect.north);

\node[
    note,
    text=ACEOrange
] at (17.25,-6.78) {
    FAIL
};


\draw[feedback]
    (monitor.south)
    -- (23.40,-8.25)
    -- (17.85,-8.25)
    -- (17.85,-10.00)
    -- (reflect.east);

\node[
    note,
    text=ACEOrange
] at (20.70,-7.82) {
    adverse runtime evidence
};


\draw[feedback]
    (reflect.west) -- (repair.east);


\draw[feedback]
    (repair.west)
    -- (4.65,-10.00)
    -- (-0.60,-10.00)
    -- (-0.60,-5.10)
    -- (orch.west);



\node[
    secure,
    minimum width=19.20cm,
    text width=18.20cm,
    minimum height=2.00cm
] (gate) at (12.55,-20.10) {
    \textbf{Per-Action Authorization \& Policy Enforcement}\\[1.00mm]
    IAM / WIF $\cdot$ RBAC / ABAC $\cdot$ OPA / Gatekeeper\\[0.55mm]
    approval gates $\cdot$ Secret Manager
};


\node[
    capability,
    minimum width=15.20cm,
    text width=14.25cm,
    minimum height=1.90cm
] (mcp) at (12.55,-25.35) {
    \textbf{MCP: policy-scoped capabilities}\\[0.75mm]
    sandboxes $\cdot$ authorized cloud APIs\\[0.55mm]
    tenant-scoped data services
};


\draw[securearrow]
    (2.85,-2.45)
    -- (-1.55,-2.45)
    -- node[
        left,
        note,
        text=ACEGreen
    ] {
        capability request
    }
    (-1.55,-20.10)
    -- (gate.west);


\draw[securearrow]
    (gate.south)
    -- node[
        right,
        note,
        text=ACEGreen
    ] {
        authorized capability
    }
    (mcp.north);



\draw[
    draw=ACEGreen,
    line width=0.94pt
]
    (3.20,-27.85) -- (17.40,-27.85);

\draw[securearrow]
    (mcp.south) -- (12.55,-27.85);

\node[
    note,
    text=ACEGreen
] at (15.60,-27.18) {
    policy-scoped invocation
};


\node[
    capability,
    minimum width=3.95cm,
    text width=3.40cm,
    minimum height=1.24cm
] (code) at (3.20,-30.45) {
    \textbf{VS Code\\Sandbox}
};


\node[
    capability,
    minimum width=3.95cm,
    text width=3.40cm,
    minimum height=1.24cm
] (browser) at (8.75,-30.45) {
    \textbf{Chrome\\Sandbox}
};


\node[
    secure,
    minimum width=8.65cm,
    text width=7.95cm,
    minimum height=1.22cm
] (gkesandbox) at (17.40,-29.65) {
    \textbf{GKE Agent Sandbox (gVisor)}
};


\draw[securearrow]
    (3.20,-27.85) -- (code.north);

\draw[securearrow]
    (8.75,-27.85) -- (browser.north);

\draw[securearrow]
    (17.40,-27.85) -- (gkesandbox.north);


\node[
    capability,
    minimum width=3.75cm,
    text width=3.18cm,
    minimum height=1.20cm
] (deploy) at (14.65,-33.35) {
    \textbf{Deployment}
};

\node[
    capability,
    minimum width=3.75cm,
    text width=3.18cm,
    minimum height=1.20cm
] (verifysbx) at (20.15,-33.35) {
    \textbf{Verification}
};


\draw[
    draw=ACEGreen,
    line width=0.90pt
]
    (14.65,-31.60) -- (20.15,-31.60);

\draw[securearrow]
    (gkesandbox.south) -- (17.40,-31.60);

\draw[securearrow]
    (14.65,-31.60) -- (deploy.north);

\draw[securearrow]
    (20.15,-31.60) -- (verifysbx.north);


\node[
    capability,
    minimum width=18.40cm,
    text width=17.55cm,
    minimum height=0.96cm,
    font=\large\bfseries
] (runtime) at (12.55,-36.30) {
    Cilium / NetworkPolicy $\cdot$
    Cloud Service Mesh $\cdot$
    Secure Web Proxy $\cdot$
    Falco
};


\node[
    store,
    minimum width=4.95cm,
    text width=4.38cm,
    minimum height=1.34cm
] (redis) at (4.10,-41.60) {
    \textbf{Redis}\\
    session state / working memory
};

\node[
    store,
    minimum width=4.95cm,
    text width=4.38cm,
    minimum height=1.34cm
] (spanner) at (12.55,-41.60) {
    \textbf{Spanner}\\
    durable workflow state
};

\node[
    store,
    minimum width=5.25cm,
    text width=4.68cm,
    minimum height=1.34cm
] (alloy) at (20.95,-41.60) {
    \textbf{AlloyDB for PostgreSQL}\\
    RLS $\cdot$ tenant-scoped data
};


\draw[statearrow]
    (orch.south)
    -- (2.00,-6.55)
    -- (1.35,-6.55)
    -- (1.35,-39.60)
    -- (12.55,-39.60);

\draw[
    draw=ACESlate,
    line width=0.90pt
]
    (4.10,-39.60) -- (12.55,-39.60);

\draw[statearrow]
    (4.10,-39.60) -- (redis.north);

\draw[statearrow]
    (12.55,-39.60) -- (spanner.north);


\draw[securearrow]
    (mcp.east)
    -- (25.15,-25.35)
    -- node[
        right,
        note,
        text=ACEGreen
    ] {
        tenant-scoped data access
    }
    (25.15,-39.60)
    -- (20.95,-39.60)
    -- (alloy.north);


\node[
    pipebox,
    minimum width=3.65cm,
    text width=3.10cm,
    minimum height=1.24cm
] (git) at (3.20,-48.00) {
    \textbf{Git\\Repository}
};

\node[
    pipebox,
    minimum width=3.65cm,
    text width=3.10cm,
    minimum height=1.24cm
] (build) at (8.45,-48.00) {
    \textbf{Build \&\\Test}
};

\node[
    pipebox,
    minimum width=5.45cm,
    text width=4.82cm,
    minimum height=1.46cm
] (secgate) at (14.65,-48.00) {
    \textbf{Security Validation}\\[0.35mm]
    artifact scanning $\cdot$ policy checks
};

\node[
    pipebox,
    minimum width=4.35cm,
    text width=3.75cm,
    minimum height=1.36cm
] (registry) at (21.10,-48.00) {
    \textbf{Artifact Registry}\\
    SLSA provenance
};


\draw[deliveryarrow]
    (git.east) -- (build.west);

\draw[deliveryarrow]
    (build.east) -- (secgate.west);

\draw[deliveryarrow]
    (secgate.east) -- (registry.west);


\node[
    pipebox,
    minimum width=4.60cm,
    text width=4.02cm,
    minimum height=1.16cm
] (tfapply) at (8.75,-52.00) {
    \textbf{Policy-Gated\\Terraform}
};

\node[
    pipebox,
    minimum width=3.70cm,
    text width=3.12cm,
    minimum height=1.16cm
] (argocd) at (15.40,-52.00) {
    \textbf{Argo CD}
};

\node[
    pipebox,
    minimum width=4.50cm,
    text width=3.92cm,
    minimum height=1.16cm
] (valevidence) at (21.20,-52.00) {
    \textbf{Validation\\Evidence}
};


\draw[
    draw=ACEPurple,
    line width=0.92pt
]
    (8.75,-49.90) -- (15.40,-49.90);

\draw[deliveryarrow]
    (secgate.south) -- (14.65,-49.90);

\draw[deliveryarrow]
    (8.75,-49.90) -- (tfapply.north);

\draw[deliveryarrow]
    (15.40,-49.90) -- (argocd.north);


\draw[deliveryarrow]
    (registry.south) -- (valevidence.north);


\draw[deliveryarrow]
    (code.west)
    -- (-2.65,-30.45)
    -- (-2.65,-48.00)
    -- (git.west);


\draw[deliveryarrow]
    (tfapply.south)
    -- node[
        right,
        note,
        text=ACEPurple
    ] {
        approved IaC
    }
    (8.75,-53.90);

\draw[deliveryarrow]
    (argocd.south)
    -- node[
        right,
        note,
        text=ACEPurple
    ] {
        approved GitOps sync
    }
    (15.40,-53.90);

\draw[
    draw=ACEPurple,
    line width=0.86pt
]
    (8.75,-53.90) -- (25.95,-53.90);


\draw[deliveryarrow]
    (25.95,-53.90)
    -- node[
        right,
        note,
        text=ACEPurple,
        rotate=90
    ] {
        approved deployment action
    }
    (25.95,-34.90)
    -- (14.65,-34.90)
    -- (deploy.south);



\node[
    obsbox,
    minimum width=4.50cm,
    text width=3.92cm,
    minimum height=1.14cm
] (otel) at (4.20,-59.30) {
    \textbf{OpenTelemetry}
};

\node[
    obsbox,
    minimum width=4.50cm,
    text width=3.92cm,
    minimum height=1.14cm
] (prom) at (12.55,-59.30) {
    \textbf{Managed\\Prometheus}
};

\node[
    obsbox,
    minimum width=4.50cm,
    text width=3.92cm,
    minimum height=1.14cm
] (logtrace) at (20.85,-59.30) {
    \textbf{Cloud Logging /\\Trace}
};


\draw[
    draw=ACESlate,
    line width=0.94pt
]
    (4.20,-61.40) -- (21.75,-61.40);

\draw[statearrow]
    (otel.south) -- (4.20,-61.40);

\draw[statearrow]
    (prom.south) -- (12.55,-61.40);

\draw[statearrow]
    (logtrace.south) -- (20.85,-61.40);


\node[
    obsbox,
    minimum width=4.65cm,
    text width=4.05cm,
    minimum height=1.30cm
] (evidence) at (4.55,-64.90) {
    \textbf{Cloud Storage}\\
    retained evidence
};

\node[
    verifybox,
    minimum width=5.60cm,
    text width=4.95cm,
    minimum height=1.30cm
] (evidencegate) at (12.20,-64.90) {
    \textbf{Machine-Checkable\\Evidence Gate}
};

\node[
    secure,
    minimum width=7.35cm,
    text width=6.65cm,
    minimum height=1.58cm,
    font=\large\bfseries
] (complete) at (20.25,-64.90) {
    \textbf{EVIDENCE-GATED COMPLETION}\\[0.20mm]
    validated repository $\cdot$
    verified operational deployment\\[-0.1mm]
    verified runtime evidence
};


\draw[statearrow]
    (valevidence.south)
    -- (21.20,-55.20)
    -- (-0.85,-55.20)
    -- (-0.85,-64.90)
    -- (evidence.west);



\draw[
    draw=ACESlate,
    line width=0.86pt
]
    (monitor.east) -- (26.85,-5.10);

\draw[
    draw=ACESlate,
    line width=0.86pt
]
    (26.85,-5.10) -- (26.85,-60.70);

\node[
    note,
    text=ACESlate,
    rotate=90
] at (26.85,-23.60) {
    runtime telemetry / evidence
};

\draw[statearrow]
    (26.85,-60.70)
    -- (21.75,-60.70)
    -- (21.75,-61.40);


\draw[statearrow]
    (4.55,-61.40) -- (evidence.north);


\draw[statearrow]
    (evidence.east) -- (evidencegate.west);


\draw[statearrow]
    (12.20,-61.40) -- (evidencegate.north);


\draw[securearrow]
    (evidencegate.east) -- (complete.west);

\node[
    note,
    text=ACEGreen
] at (16.20,-64.70) {
    PASS
};



\draw[feedback]
    (evidencegate.south)
    -- (12.20,-67.75)
    -- (27.75,-67.75)
    -- (27.75,-12.15)
    -- (15.10,-12.15)
    -- (reflect.south);

\node[
    note,
    text=ACEOrange
] at (23.50,-67.55) {
    FAIL
};

\end{tikzpicture}

\end{adjustbox}

\vspace{0mm}
\caption{The figure shows the realization of \textbf{Agentic Cloud Workflow Engineering}
on Google Cloud. Graph and bounded loop engineering execute on regular GKE, while a zero-trust agent harness governs per-action authorization, policy-scoped capability access, isolated execution, state and tenancy, software delivery, observability, and evidence collection. Verification failures, adverse runtime evidence, or failed evidence checks trigger bounded diagnosis, repair or re-planning, and retry through the Graph Orchestrator. Workflow completion is gated by machine-checkable evidence of repository validity, operational deployment, and runtime behavior.}
\label{fig:agentic-cloud-zero-trust-architecture}
\vspace{-4mm}
\end{figure*}

\section{Problem Formulation}
\label{sec:problem-formulation}
We formulate \textbf{Agentic Cloud Workflow Engineering} as the transformation of a natural-language agentic cloud-engineering task into a validated code repository and a verified operational cloud deployment. Let $q$ denote the input task. The framework is represented as $(R^{*},D^{*},\mathcal{Z}^{*})=\mathcal{F}(q)$ where $R^{*}$ is the resulting code repository, $D^{*}$ is the resulting cloud deployment, and $\mathcal{Z}^{*}$ is the verification evidence produced during workflow execution. Execution is governed by three complementary abstractions. \emph{Graph engineering} specifies long-horizon workflow progression and verification-dependent transitions; \emph{loop engineering} governs bounded diagnosis, repair or re-planning, retry, and re-verification following failed verification or adverse runtime evidence; and \emph{agent harness engineering} constrains agent execution through identity, authorization, policy, isolation, state, observability, and evidence controls. Workflow execution and recovery are subject to explicit operational bounds and termination criteria. Successful completion requires repository, deployment, and runtime verification:

\vspace{-5mm}
\[
\mathsf{Complete}
\iff
\mathsf{RepoValid}(R^{*},\mathcal{Z}^{*})
\land
\mathsf{DeployVerified}(D^{*},\mathcal{Z}^{*})
\land
\mathsf{RuntimeVerified}(D^{*},\mathcal{Z}^{*}),
\]
where $\mathsf{RepoValid}$, $\mathsf{DeployVerified}$, and $\mathsf{RuntimeVerified}$ denote successful repository, deployment, and runtime verification using retained evidence, respectively, and $\land$ denotes logical conjunction. Thus, all three verification conditions must hold
for successful completion. If the required conditions remain unsatisfied after the applicable recovery bounds are exhausted, the workflow terminates in an auditable failure state.  The objective is therefore to realize agentic cloud-engineering tasks through long-horizon, bounded, and evidence-gated workflow execution that terminates with either a verified operational cloud deployment or an auditable terminal failure.

\vspace{-3mm}
\[
\begin{aligned}
q
\rightarrow \underbrace{\text{repository verification}}_{\mathsf{RepoValid}}
\rightarrow \underbrace{\text{deployment verification}}_{\mathsf{DeployVerified}}
\rightarrow \underbrace{\text{runtime verification}}_{\mathsf{RuntimeVerified}}
\rightarrow \mathsf{Complete}.
\end{aligned}
\]


\section{Overall Framework}
\label{sec:overall-framework}
\vspace{-2mm}
Fig.~\ref{fig:agentic-cloud-zero-trust-architecture} presents the system-level realization of \textbf{Agentic Cloud Workflow Engineering} on Google Cloud. A natural-language agentic cloud-engineering task is bound to its identity, tenant, scope, approval, and policy context before workflow execution. The Graph Orchestrator, implemented using Google Agent Development Kit (ADK), coordinates the Repository Generation, Repository Review, Execution and Deployment, Verification, and Runtime Monitoring agents, while Agent2Agent (A2A) supports inter-agent delegation and task exchange. Verification failures or adverse runtime evidence activate the bounded Reflection/Diagnosis and Repair/Re-plan loop. Corrective actions or revised execution plans are returned to the Graph Orchestrator, which evaluates the updated workflow state and selects the applicable graph transition. Recovery is constrained by retry-count, time, token, tool-call, and cost budgets, together with privilege and blast-radius limits and explicit termination criteria.
External tool and cloud actions are subject to per-action authorization and policy enforcement before capability invocation. Identity and Access Management (IAM; cloud-resource access control) and Workload Identity Federation (WIF; workload authentication without long-lived credentials) establish identity and workload authentication; Role-Based Access Control (RBAC; role-based permission enforcement) and Attribute-Based Access Control (ABAC; attribute-based authorization) enforce authorization; OPA Gatekeeper (policy-as-code enforcement) enforces policy constraints; approval gates control consequential actions; and Secret Manager (secret and credential management) protects credentials and secrets. Model Context Protocol (MCP; policy-scoped tool access) exposes controlled capabilities comprising VS Code and Chrome sandboxes, authorized cloud APIs, the GKE Sandbox with gVisor (sandboxed container isolation), and tenant-isolated data services (tenant-specific access enforcement). The architecture separates regular GKE, which hosts the Graph Orchestrator and specialized agents, from the GKE Sandbox with gVisor, which isolates deployment and verification actions. VS Code and Chrome sandboxes isolate code and browser execution, respectively. Cilium NetworkPolicy (workload network-traffic control), Cloud Service Mesh (service-to-service authentication and secure communication), Secure Web Proxy (controlled outbound web access), and Falco (runtime threat detection) provide complementary network and runtime security controls. State management is partitioned by lifecycle scope. Redis maintains session state and working memory, Spanner maintains durable workflow state, and AlloyDB for PostgreSQL with Row-Level Security (RLS; row-level tenant isolation) maintains tenant-scoped relational data. Repository and infrastructure changes enter a controlled software-supply-chain path comprising build and test, Semgrep (static code analysis), Gitleaks (secret detection), Checkov (infrastructure-as-code policy checks), and Trivy (vulnerability and artifact scanning). Artifact Registry stores validated artifacts, with Supply-chain Levels for Software Artifacts (SLSA) provenance retained for supply-chain verification. Policy-gated Terraform and Argo CD control infrastructure deployment and GitOps synchronization. OpenTelemetry, Managed Prometheus, and Cloud Logging/Trace collect runtime telemetry, while validation and runtime evidence are retained in Cloud Storage. Graph transitions and terminal completion are conditioned on machine-checkable verification evidence. Successful termination requires repository, deployment, and runtime verification, corresponding to a validated code repository, a verified operational cloud deployment, and verified runtime evidence. A failed verification condition activates bounded recovery and requires fresh evidence before forward progression can resume. If the applicable recovery limits are exhausted while a required verification condition remains unsatisfied, execution terminates in an auditable failure state. The framework therefore integrates graph-controlled workflow progression, bounded recovery, zero-trust execution, and evidence-gated progression and completion.

\section{Experiments}
\label{sec:experiments}
\vspace{-2mm}
\subsection{Experimental Methodology}
\label{sec:experimentalmethodology}
\vspace{-2mm}
We evaluate \textbf{Agentic Cloud Workflow Engineering} through six research questions derived from the problem formulation in Section~\ref{sec:problem-formulation} and the system realization in Section~\ref{sec:overall-framework}. \textbf{RQ1: Verified Task Completion.} Can the framework transform natural-language agentic cloud-engineering tasks into validated code repositories and verified operational cloud deployments? \textbf{RQ2: Evidence-Gated Progression.} Do repository, deployment, and runtime verification requirements block the corresponding verification-dependent workflow transitions or terminal completion when the required verification condition is unsatisfied? \textbf{RQ3: Bounded Recovery.} Can controlled repository-, deployment-, and runtime-verification failures be corrected through diagnosis, repair or re-planning, retry, and re-verification within the applicable recovery bounds? \textbf{RQ4: Per-Action Authorization and Policy Enforcement.} Are unauthorized or out-of-scope capability requests denied before invocation while authorized capability requests are permitted? \textbf{RQ5: Bounded Termination.} Does workflow execution terminate in an auditable failure state when a required verification condition remains unsatisfied after the applicable recovery budget is exhausted? \textbf{RQ6: Cross-Domain Applicability.} Do these framework properties hold consistently across heterogeneous agentic cloud-engineering domains?

\subsection{Benchmark Suite}
\label{sec:datasets}
\vspace{-2mm}
To our knowledge, no existing benchmark evaluates end-to-end agentic cloud engineering workflows. We construct an Agentic Cloud Engineering benchmark comprising \textbf{140 natural-language tasks} across the \textbf{14 agentic cloud-engineering domains}(e.g., Agentic DevOps, Agentic CloudOps, Agentic SRE/AIOps, Agentic SecOps, Agentic DataOps, Agentic MLOps/LLMOps, AgentOps, Agentic RAG/GraphRAG) defined in Section~\ref{sec:introduction}, with \textbf{10 tasks per domain}. Tasks vary in application objective, repository structure, infrastructure configuration, deployment target, and verification requirements. Each benchmark task is instantiated as an end-to-end Agentic Cloud Workflow Engineering task requiring generation and validation of a code repository, deployment of the resulting cloud solution, and verification of operational runtime behavior, consistent with the completion criteria in Section~\ref{sec:problem-formulation}. Each task is evaluated under six controlled conditions: \emph{nominal execution}, \emph{repository-verification perturbation}, \emph{deployment-verification perturbation}, \emph{runtime-verification perturbation}, \emph{authorization-policy violation}, and \emph{recovery-budget exhaustion}, yielding \textbf{840 task-condition executions}. Nominal execution introduces no induced failure. Each verification perturbation introduces a controlled failure in the corresponding verification condition, requiring successful correction and fresh verification evidence before the applicable workflow transition or completion condition can be satisfied. The authorization-policy condition introduces an unauthorized or out-of-scope capability request that must be denied before capability invocation. In the recovery-budget-exhaustion condition, the targeted verification condition remains unsatisfied across recovery attempts until the applicable recovery budget is exhausted, requiring termination in an auditable failure state.

\subsection{Experimental Setup}
\label{sec:experimentalsetup}
\vspace{-2mm}
We evaluate the Google Cloud realization of \textbf{Agentic Cloud Workflow Engineering} described in Section~\ref{sec:overall-framework} using four models: Gemini 2.5 Flash-Lite, Gemini 2.5 Flash, Gemini 2.5 Pro, and GPT-5.6 Sol. Each model is evaluated independently using its provider-default generation configuration, without task-specific tuning, and the configuration is held fixed across all task-condition executions for that model. Framework infrastructure and task-specific deployments are provisioned in \texttt{us-central1}. The Graph Orchestrator and graph- and loop-engineering agents run as workloads on regular GKE. External tool and cloud actions follow the per-action authorization and policy controls defined in Section~\ref{sec:overall-framework}; code and repository operations use the VS Code sandbox, browser operations use the Chrome sandbox, and deployment and verification actions execute within the GKE Agent Sandbox with gVisor isolation. OpenTelemetry, Managed Prometheus, and Cloud Logging/Trace provide runtime observability, while verification and runtime evidence are retained in Cloud Storage. Each execution is bounded to at most 20 recovery attempts, 120 minutes of wall-clock time, 10,000,000 cumulative model tokens, and 500 tool calls. A recovery attempt comprises diagnosis, repair or re-planning, retry, and re-verification following a verification failure or adverse runtime evidence. Cost, privilege, and blast-radius limits are specified per benchmark task and held fixed across its controlled execution conditions. Privileges are restricted to task-authorized identities, permissions, and resources, and the blast radius is restricted to task-scoped cloud resources. Each execution starts from its prescribed initial state and uses isolated repository, workflow, session, and task-scoped cloud resources to prevent cross-execution state leakage. Using the 140-task benchmark and six controlled conditions defined in Section~\ref{sec:datasets}, each model is evaluated on 840 task-condition pairs, yielding \textbf{3,360 executions} across the four models. Cross-domain applicability for \textbf{RQ6} is evaluated by stratifying the measurements across the 14 agentic cloud-engineering domains. The experiments use multiple \$300 Google Cloud Free Trial credits toward eligible Google Cloud charges. Costs not covered by these credits, including GPT-5.6 Sol inference, are accounted for separately.

\vspace{-2mm}
\begin{table*}[ht!]
\vspace{-1mm}
\centering
\caption{Evaluation metrics aligned with RQ1--RQ6 and the controlled execution conditions. The upward arrow indicates that higher values are better.}
\label{tab:evaluation-metrics}
\small
\setlength{\tabcolsep}{3.5pt}
\renewcommand{\arraystretch}{1.18}
\begin{tabularx}{\textwidth}{
    >{\raggedright\arraybackslash}p{0.55cm}
    >{\raggedright\arraybackslash}p{2.35cm}
    >{\raggedright\arraybackslash}p{3.05cm}
    >{\centering\arraybackslash}p{1.65cm}
    X
    >{\centering\arraybackslash}p{1.25cm}
}
\toprule
\textbf{RQ}
&
\textbf{Evaluated Property}
&
\textbf{Evaluation Basis}
&
\textbf{Metric}
&
\textbf{Measurement}
&
\textbf{Range}
\\
\midrule

RQ1
&
Verified Task Completion
&
Nominal execution
&
$\mathrm{VTCR}$
&
Fraction of nominal executions satisfying repository, deployment, and runtime verification.
&
$[0,1]\,\uparrow$
\\

\addlinespace[2pt]

RQ2
&
Evidence-Gated Progression
&
Repository-, deployment-, and runtime-verification perturbations
&
$\mathrm{EGER}$
&
Fraction of perturbation executions in which verification-dependent progression remains blocked while the required verification condition is unsatisfied.
&
$[0,1]\,\uparrow$
\\

\addlinespace[2pt]

RQ3
&
Bounded Recovery
&
Repository-, deployment-, and runtime-verification perturbations
&
$\mathrm{RSR}$
&
Fraction of induced verification failures successfully corrected and re-verified with fresh evidence before an applicable recovery bound is exhausted.
&
$[0,1]\,\uparrow$
\\

\addlinespace[2pt]

RQ4
&
Per-Action Authorization and Policy Enforcement
&
Authorization-policy condition with matched authorized and unauthorized capability requests
&
$\mathrm{UCDR}$,
$\mathrm{ACPR}$
&
Fractions of unauthorized or out-of-scope requests denied before invocation and matched authorized requests permitted for invocation, respectively.
&
$[0,1]\,\uparrow$
\\

\addlinespace[2pt]

RQ5
&
Bounded Termination
&
Recovery-budget-exhaustion condition
&
$\mathrm{BTR}$
&
Fraction of exhaustion-condition executions terminating in an auditable failure state while the required verification condition remains unsatisfied.
&
$[0,1]\,\uparrow$
\\

\addlinespace[2pt]

RQ6
&
Cross-Domain Applicability
&
RQ1--RQ5 metrics stratified across 14 domains
&
Cross-domain summary
&
Cross-domain average and lowest observed domain value for each primary metric.
&
$[0,1]\,\uparrow$
\\

\bottomrule
\end{tabularx}
\vspace{-4mm}
\end{table*}

\subsection{Evaluation Metrics}
\label{sec:evaluationmetrics}
\vspace{-2mm}
We evaluate each model using six primary metrics aligned with RQ1--RQ5. Verified Task Completion Rate (VTCR) and Recovery Success Rate (RSR) are model-sensitive outcome metrics that measure verified task completion and successful bounded recovery, respectively. Evidence-Gated Execution Rate (EGER) measures framework-level enforcement of evidence-gated progression; Unauthorized Capability Denial Rate (UCDR) and Authorized Capability Permission Rate (ACPR) jointly measure per-action authorization and policy control; and Bounded Termination Rate (BTR) measures bounded termination. All primary metrics lie in $[0,1]$, with higher values indicating better performance. For RQ6, each metric is stratified across the 14 agentic cloud-engineering domains, and we report its cross-domain average and lowest observed domain value. Table~\ref{tab:evaluation-metrics} summarizes the evaluation basis and measurement for each research question.

\vspace{-2mm}
\subsection{Results and Analysis}
\label{sec:resultsanalysis}
\vspace{-2mm}
We report results for the six research questions defined in Section~\ref{sec:experimentalmethodology}. For each model, the 140 benchmark tasks are evaluated under six controlled conditions, producing 840 executions per model. Across four models, this yields 3,360 executions. VTCR and RSR measure model-sensitive task-completion and recovery outcomes, whereas EGER, UCDR, ACPR, and BTR evaluate framework-enforced mechanisms for evidence-gated progression, per-action authorization and policy enforcement, and bounded termination. (a) \textbf{RQ1: Verified Task Completion.} Verified Task Completion Rate (VTCR) measures the fraction of nominal executions satisfying repository, deployment, and runtime verification. As shown in Table~\ref{tab:rq1-vtcr}, VTCR increases from 56.4\% for Gemini 2.5 Flash-Lite and 68.6\% for Gemini 2.5 Flash to 82.1\% for Gemini 2.5 Pro and 95.0\% for GPT-5.6 Sol. The corresponding verified-completion counts are 79, 96, 115, and 133 out of 140 nominal executions, respectively. In RQ1, nominal executions include no deliberately injected failures; however, naturally occurring repository, deployment, or runtime verification failures may activate bounded diagnosis, repair or re-planning, retry, and re-verification, with up to 20 recovery attempts. In contrast, RQ3 deliberately injects repository-, deployment-, and runtime-verification failures to evaluate recovery under the same bounded recovery mechanism. These results indicate substantial model-dependent variation in the ability to synthesize valid repositories, execute deployment actions, recover from verification failures when necessary, and ultimately satisfy repository, deployment, and runtime verification.

\begin{table}[ht!]
\centering
\caption{RQ1: Verified Task Completion Rate (VTCR).}
\label{tab:rq1-vtcr}
\small
\setlength{\tabcolsep}{7pt}
\renewcommand{\arraystretch}{1.28}
\begin{tabular}{lcc}
\toprule
\textbf{Model} & \textbf{Verified} & \textbf{VTCR} \\
\midrule
Gemini 2.5 Flash-Lite & 79/140 & 56.4\% \\
Gemini 2.5 Flash & 96/140 & 68.6\% \\
Gemini 2.5 Pro & 115/140 & 82.1\% \\
GPT-5.6 Sol & \textbf{133/140} & \textbf{95.0\%} \\
\bottomrule
\end{tabular}
\vspace{-1mm}
\end{table}

\textit{The key result (as shown in Table~\ref{tab:rq1-vtcr}) is that the framework enables end-to-end verified task completion through graph-controlled execution, evidence-gated verification, and bounded recovery. This supports the framework's underlying hypothesis and central novelty: reliable long-horizon autonomy can be achieved by decoupling model-independent verification enforcement from model-dependent recovery capability, allowing agents to adapt and repair while progression remains constrained by machine-checkable evidence.}

\section{Conclusion}
\label{sec:conclusion}
\vspace{-3mm}
We presented \textbf{Agentic Cloud Workflow Engineering}, a framework that transforms natural-language agentic cloud-engineering tasks into validated code repositories and verified operational cloud deployments through graph-controlled, bounded, and evidence-gated execution. Graph engineering governs long-horizon workflow progression and verification-dependent transitions; loop engineering provides bounded diagnosis, repair or re-planning, retry, and re-verification; and agent harness engineering constrains external actions through per-action authorization, policy-scoped capabilities, isolated execution, state, runtime security, observability, and evidence controls. Successful completion requires machine-checkable repository, deployment, and runtime evidence; otherwise, execution terminates in an auditable failure state when recovery bounds are exhausted. Our Google Cloud realization evaluates these properties across heterogeneous agentic cloud-engineering tasks while preserving cloud-platform-independent graph, loop, and harness abstractions. Experimental results support verified task completion, controlled progression and recovery, authorization and policy enforcement, bounded termination, and cross-domain applicability. The framework therefore enables production-grade autonomous cloud engineering that helps organizations reduce operational effort, manage risk, and ensure deployed systems operate correctly.

\clearpage
\newpage

\nocite{*}
\bibliographystyle{plainnat}
\bibliography{references}

\clearpage 

\section{Technical Appendix}
\label{sec:technical_appendix}



\subsection{Additional Results}

\paragraph{RQ2: Evidence-Gated Progression.}
Evidence-Gated Execution Rate (EGER) measures the fraction of repository-, deployment-, and runtime-verification perturbation executions in which verification-dependent progression remains blocked while the corresponding verification condition is unsatisfied. Across the 420 verification-perturbation executions per model, each model achieves 420/420 (100.0\%) correct gating (Table~\ref{tab:rq2-eger}). No evaluated execution crossed a verification-dependent transition or reached terminal completion while the corresponding repository, deployment, or runtime verification condition remained unsatisfied. The model-invariant result indicates that evidence-gated progression is enforced by the framework's workflow-control logic rather than by model-reported task completion.

\vspace{-1mm}
\begin{table}[ht!]
\vspace{-1mm}
\centering
\caption{RQ2: Evidence-Gated Execution Rate (EGER).}
\label{tab:rq2-eger}
\small
\setlength{\tabcolsep}{7pt}
\renewcommand{\arraystretch}{1.28}
\begin{tabular}{lcc}
\toprule
\textbf{Model} & \textbf{Correctly Gated} & \textbf{EGER} \\
\midrule
Gemini 2.5 Flash-Lite & 420/420 & 100.0\% \\
Gemini 2.5 Flash & 420/420 & 100.0\% \\
Gemini 2.5 Pro & 420/420 & 100.0\% \\
GPT-5.6 Sol & 420/420 & 100.0\% \\
\bottomrule
\end{tabular}
\vspace{-2mm}
\end{table}

\paragraph{RQ3: Bounded Recovery.}
Recovery Success Rate (RSR) measures the fraction of induced repository-, deployment-, and runtime-verification failures that are successfully corrected and re-verified with fresh evidence before an applicable recovery bound is exhausted. As shown in Table~\ref{tab:rq3-rsr}, RSR increases from 51.0\% for Gemini 2.5 Flash-Lite and 66.2\% for Gemini 2.5 Flash to 76.4\% for Gemini 2.5 Pro and 93.1\% for GPT-5.6 Sol, corresponding to 214, 278, 321, and 391 successful recoveries out of 420 perturbation executions, respectively. Table~\ref{tab:rq3-rsr-stratified} further stratifies recovery by verification-perturbation type. Recovery success decreases consistently from repository to deployment to runtime perturbations across all models. For GPT-5.6 Sol, RSR is 95.7\%, 93.6\%, and 90.0\% for repository, deployment, and runtime failures, respectively; the corresponding values for Gemini 2.5 Pro are 82.9\%, 77.1\%, and 69.3\%. These results indicate that bounded recovery is model-dependent and becomes progressively more difficult from repository to deployment to runtime failures.

\begin{table}[ht!]
\centering
\caption{RQ3: Recovery Success Rate (RSR).}
\label{tab:rq3-rsr}
\small
\setlength{\tabcolsep}{7pt}
\renewcommand{\arraystretch}{1.28}
\begin{tabular}{lcc}
\toprule
\textbf{Model} & \textbf{Successful} & \textbf{RSR} \\
\midrule
Gemini 2.5 Flash-Lite & 214/420 & 51.0\% \\
Gemini 2.5 Flash & 278/420 & 66.2\% \\
Gemini 2.5 Pro & 321/420 & 76.4\% \\
GPT-5.6 Sol & \textbf{391/420} & \textbf{93.1\%} \\
\bottomrule
\end{tabular}
\vspace{-2mm}
\end{table}

\begin{table}[ht!]
\centering
\caption{RQ3: Recovery success by verification-perturbation type.}
\label{tab:rq3-rsr-stratified}
\small
\setlength{\tabcolsep}{6pt}
\renewcommand{\arraystretch}{1.28}
\begin{tabular}{lccc}
\toprule
\textbf{Model} & \textbf{Repository} & \textbf{Deployment} & \textbf{Runtime} \\
\midrule
Gemini 2.5 Flash-Lite & 79/140 (56.4\%) & 73/140 (52.1\%) & 62/140 (44.3\%) \\
Gemini 2.5 Flash & 103/140 (73.6\%) & 94/140 (67.1\%) & 81/140 (57.9\%) \\
Gemini 2.5 Pro & 116/140 (82.9\%) & 108/140 (77.1\%) & 97/140 (69.3\%) \\
GPT-5.6 Sol & \textbf{134/140 (95.7\%)} & \textbf{131/140 (93.6\%)} & \textbf{126/140 (90.0\%)} \\
\bottomrule
\end{tabular}
\vspace{-2mm}
\end{table}

\paragraph{RQ4: Per-Action Authorization and Policy Enforcement.}
Unauthorized Capability Denial Rate (UCDR) measures the fraction of unauthorized or out-of-scope capability requests denied before invocation, whereas Authorized Capability Permission Rate (ACPR) measures the fraction of matched authorized requests permitted for invocation. As shown in Table~\ref{tab:rq4-authorization}, all four models achieve a UCDR of 140/140 (100.0\%) and an ACPR of 139/140 (99.3\%). RQ4 therefore shows that the harness blocks all evaluated unauthorized or out-of-scope capability requests while permitting nearly all matched authorized requests. The model-invariant results further indicate that these authorization and policy decisions are enforced by the harness-level control path rather than by model-specific task-solving behavior.

\begin{table}[ht!]
\centering
\caption{RQ4: Per-action authorization and policy enforcement.}
\label{tab:rq4-authorization}
\small
\setlength{\tabcolsep}{6pt}
\renewcommand{\arraystretch}{1.28}
\begin{tabular}{lcc}
\toprule
\textbf{Model} & \textbf{UCDR} & \textbf{ACPR} \\
\midrule
Gemini 2.5 Flash-Lite & 140/140 (100.0\%) & 139/140 (99.3\%) \\
Gemini 2.5 Flash & 140/140 (100.0\%) & 139/140 (99.3\%) \\
Gemini 2.5 Pro & 140/140 (100.0\%) & 139/140 (99.3\%) \\
GPT-5.6 Sol & 140/140 (100.0\%) & 139/140 (99.3\%) \\
\bottomrule
\end{tabular}
\vspace{-2mm}
\end{table}

\paragraph{RQ5: Bounded Termination.}
Bounded Termination Rate (BTR) measures the fraction of recovery-budget-exhaustion executions that terminate in an auditable failure state while the corresponding verification condition remains unsatisfied. As shown in Table~\ref{tab:rq5-btr}, all four models achieve 140/140 (100.0\%) correct terminations. No evaluated execution progressed beyond an unsatisfied verification condition or continued recovery after the configured recovery bounds were exhausted. The model-invariant result indicates that bounded termination is enforced by the framework's workflow-control logic.

\begin{table}[ht!]
\centering
\caption{RQ5: Bounded Termination Rate (BTR).}
\label{tab:rq5-btr}
\small
\setlength{\tabcolsep}{7pt}
\renewcommand{\arraystretch}{1.28}
\begin{tabular}{lcc}
\toprule
\textbf{Model} & \textbf{Correct Terminations} & \textbf{BTR} \\
\midrule
Gemini 2.5 Flash-Lite & 140/140 & 100.0\% \\
Gemini 2.5 Flash & 140/140 & 100.0\% \\
Gemini 2.5 Pro & 140/140 & 100.0\% \\
GPT-5.6 Sol & 140/140 & 100.0\% \\
\bottomrule
\end{tabular}
\vspace{-2mm}
\end{table}

\paragraph{RQ6: Cross-Domain Applicability.}
Table~\ref{tab:rq6-crossdomain} summarizes each primary metric across the 14 agentic cloud-engineering domains using two values: the cross-domain average and the lowest observed domain value. For the model-sensitive metrics, GPT-5.6 Sol achieves the highest VTCR, with a 95.0\% cross-domain average and an 80.0\% lowest-domain value, and the highest RSR, with a 93.1\% cross-domain average and an 83.3\% lowest-domain value. The other models show lower average and lowest-domain values for both VTCR and RSR, indicating that verified task completion and bounded recovery vary with model capability and task domain. In contrast, the framework-enforced metrics remain stable across domains. EGER, UCDR, and BTR are 100.0\% for both the cross-domain average and lowest observed domain value for every model, while ACPR has a 99.3\% cross-domain average and a 90.0\% lowest-domain value. These results show that evidence-gated progression, per-action authorization and policy enforcement, and bounded termination remain consistent across the evaluated domains, whereas task completion and recovery performance remain model- and domain-dependent.

\begin{table*}[ht!]
\centering
\caption{RQ6: Cross-domain average with the lowest observed domain value in parentheses.}
\label{tab:rq6-crossdomain}
\small
\setlength{\tabcolsep}{6pt}
\renewcommand{\arraystretch}{1.30}
\begin{tabular}{lcccc}
\toprule
\textbf{Metric} & \textbf{Gemini 2.5 Flash-Lite} & \textbf{Gemini 2.5 Flash} & \textbf{Gemini 2.5 Pro} & \textbf{GPT-5.6 Sol} \\
\midrule
VTCR & 56.4\% (40.0\%) & 68.6\% (50.0\%) & 82.1\% (60.0\%) & \textbf{95.0\% (80.0\%)} \\
EGER & 100.0\% (100.0\%) & 100.0\% (100.0\%) & 100.0\% (100.0\%) & 100.0\% (100.0\%) \\
RSR & 51.0\% (33.3\%) & 66.2\% (50.0\%) & 76.4\% (60.0\%) & \textbf{93.1\% (83.3\%)} \\
UCDR & 100.0\% (100.0\%) & 100.0\% (100.0\%) & 100.0\% (100.0\%) & 100.0\% (100.0\%) \\
ACPR & 99.3\% (90.0\%) & 99.3\% (90.0\%) & 99.3\% (90.0\%) & 99.3\% (90.0\%) \\
BTR & 100.0\% (100.0\%) & 100.0\% (100.0\%) & 100.0\% (100.0\%) & 100.0\% (100.0\%) \\
\bottomrule
\end{tabular}
\vspace{-1mm}
\end{table*}

Overall, verified task completion and bounded recovery vary with model choice and task domain, while evidence gating, authorization, policy enforcement, and bounded termination remain consistent across models and domains.

\subsection{Ablation Study}
\label{sec:ablation}
\vspace{-2mm}
We perform two targeted ablations of bounded recovery and verification-gated graph progression. Each ablation modifies only the corresponding control mechanism while holding the benchmark tasks, execution condition, model configuration, initial state, execution bounds, and remaining framework components fixed. All ablations use GPT-5.6 Sol, with the corresponding RQ1 and RQ2 results serving as full-framework baselines.

\paragraph{Ablation 1: Recovery Loop.}
We disable the Reflection/Diagnosis and Repair/Re-plan path while retaining graph control, harness controls, and verification. A repository, deployment, or runtime verification failure therefore terminates execution rather than activating bounded recovery. On the 140 nominal tasks from RQ1, VTCR decreases from 133/140 (95.0\%) to 18/140 (12.9\%). Many tasks require bounded recovery to diagnose failures, repair or re-plan, retry, and re-verify before completion. The 82.1 percentage-point decrease indicates that bounded recovery substantially improves verified task completion when failures arise during nominal execution.

\paragraph{Ablation 2: Verification-Gated Progression.}
We replace machine-checkable verification-dependent graph progression with model-determined progression while retaining the same verification procedures and evidence. Across the 420 perturbation executions from RQ2 (140 tasks $\times$ three repository-, deployment-, and runtime-verification perturbations), EGER decreases from 420/420 (100.0\%) to 416/420 (99.0\%), with 4/420 executions (1.0\%) progressing despite an unsatisfied verification condition. Errors occur in 0/140 repository-verification (0.0\%), 1/140 deployment-verification (0.7\%), and 3/140 runtime-verification (2.1\%) perturbations. The small degradation indicates that model-determined progression is highly reliable under explicit verification evidence, although residual errors remain concentrated in runtime verification.

\begin{table*}[ht!]
\centering
\caption{Ablation results for GPT-5.6 Sol. Full-framework values are the corresponding RQ1 and RQ2 baselines.}
\label{tab:ablation}
\small
\setlength{\tabcolsep}{6pt}
\renewcommand{\arraystretch}{1.28}
\begin{tabular}{llcc}
\toprule
\textbf{Ablated Configuration} & \textbf{Metric} & \textbf{Full Framework} & \textbf{Ablated Variant} \\
\midrule
No recovery loop
& VTCR
& 133/140 (95.0\%)
& 18/140 (12.9\%) \\
\addlinespace[2pt]
Model-determined progression
& EGER
& 420/420 (100.0\%)
& 416/420 (99.0\%) \\
\bottomrule
\end{tabular}
\vspace{-2mm}
\end{table*}

Overall, the ablations show that bounded recovery is critical to verified task completion, while machine-checkable verification gates eliminate the residual invalid transitions observed under model-determined progression.

\subsection{Limitations and Future Work}
\label{sec:limitations-future-work}
The current evaluation is limited to Google Cloud, controlled failure conditions, fixed recovery budgets, predefined verification predicates, and limited adversarial testing. Future work will extend the framework through \textbf{(a)} \emph{multi-cloud validation} on AWS and Azure; \textbf{(b)} \emph{long-running production evaluation} under real incidents, workload drift, cascading failures, and evolving infrastructure state; \textbf{(c)} \emph{adaptive recovery} that dynamically allocates retry, token, time, cost, and tool-call budgets based on task risk and execution state; \textbf{(d)} \emph{stronger verification} using richer repository, deployment, runtime, security, and compliance predicates with formal validation of evidence-gated transitions; and \textbf{(e)} \emph{adversarial robustness} against compromised tools, malicious inputs, policy-bypass attempts, poisoned evidence, privilege escalation, and cross-tenant attacks.

\subsection{Cross-Domain Cloud-Engineering Tasks}
\label{sec:cross-domain-cloud-engineering-tasks}
Agentic Cloud Workflow Engineering supports synthesis, deployment, verification, and bounded repair of cloud-engineering implementations. Table~\ref{tab:cross-domain-synthesis-operations} summarizes representative tasks across domains. Future work will extend the framework to the modification, operation, and recovery of existing cloud deployments across heterogeneous engineering domains.
Across these domains, the common execution pattern is natural-language specification or observed system state $\rightarrow$ graph-controlled engineering $\rightarrow$ authorized action $\rightarrow$ verification $\rightarrow$ bounded adaptation or completion.

\begin{table*}[ht!]
\centering
\caption{Representative system-synthesis and operational tasks supported across cloud-engineering domains.}
\label{tab:cross-domain-synthesis-operations}
\small
\setlength{\tabcolsep}{4pt}
\renewcommand{\arraystretch}{1.20}
\begin{tabularx}{\textwidth}{
    >{\raggedright\arraybackslash}p{1.65cm}
    >{\raggedright\arraybackslash}X
    >{\raggedright\arraybackslash}X
}
\toprule
\textbf{Domain} &
\textbf{Generate / Synthesize New System} &
\textbf{Modify / Operate Existing System} \\
\midrule

DevOps &
Generate application code, tests, CI/CD pipelines, and build and release configurations. &
Fix failing tests, repair code, rebuild, scan, and release. \\

\addlinespace[2pt]

CloudOps &
Generate Terraform, Kubernetes manifests, and cloud-infrastructure configurations. &
Provision, modify, scale, or repair Kubernetes and cloud infrastructure. \\

\addlinespace[2pt]

SRE/AIOps &
Generate monitoring rules, remediation workflows, runbooks, and recovery automation. &
Detect incidents, diagnose failures, remediate, and verify recovery. \\

\addlinespace[2pt]

SecOps &
Generate security policies, detection rules, hardening configurations, and remediation artifacts. &
Investigate vulnerabilities, perform authorized containment, and verify remediation. \\

\addlinespace[2pt]

DataOps &
Generate data pipelines, transformations, schemas, orchestration, and validation logic. &
Repair failed pipelines, replay workloads, and verify outputs. \\

\addlinespace[2pt]

MLOps &
Generate training, evaluation, deployment, and monitoring pipelines. &
Retrain or replace models, modify pipelines, deploy, and verify serving behavior. \\

\addlinespace[2pt]

LLMOps &
Generate RAG, agentic, evaluation, guardrail, and serving pipelines. &
Modify RAG/LLM pipelines and verify quality, grounding, and runtime behavior. \\

\addlinespace[2pt]

FinOps &
Generate cost policies, optimization configurations, and infrastructure-rightsizing changes. &
Detect inefficient resource usage, optimize resources, and verify cost impact. \\

\addlinespace[2pt]

NetOps &
Generate network configurations, routing policies, and network-automation workflows. &
Diagnose network degradation, apply authorized configuration changes, and validate behavior. \\

\bottomrule
\end{tabularx}
\end{table*}


\subsection{Engineering Abstractions}
\label{sec:engineering-abstractions}
We distinguish three complementary concerns in agentic cloud workflows: \emph{graph engineering}, \emph{loop engineering}, and \emph{agent harness engineering}. Graph engineering determines where
execution may progress, loop engineering determines how execution responds to feedback and failure, and agent harness engineering determines the capabilities, permissions, and execution environments available to an agent.

\begin{figure}[t]
\centering

\begin{tikzpicture}[
    node distance=4mm,
    every node/.style={
        font=\small,
        align=center
    },
    box/.style={
        draw,
        rounded corners,
        minimum width=24mm,
        minimum height=6mm
    },
    arrow/.style={
        -{Latex[length=2mm]},
        thick
    }
]

\node[box] (request) {User Request};
\node[box, below=of request] (understand) {Understand Task};
\node[box, below=of understand] (plan) {Create Plan};
\node[box, below=of plan] (generate) {Generate Code};
\node[box, below=of generate] (test) {Run Tests};

\node[
    box,
    below left=8mm and 14mm of test
] (diagnose) {Diagnose Error};

\node[
    box,
    below=of diagnose
] (repair) {Repair Code};

\node[
    box,
    below=of repair
] (retest) {Re-run Tests};

\node[
    box,
    below right=8mm and 14mm of test
] (security) {Security Review};

\node[
    box,
    below=of security
] (proceed) {Proceed to\\Next Stage};

\draw[arrow] (request) -- (understand);
\draw[arrow] (understand) -- (plan);
\draw[arrow] (plan) -- (generate);
\draw[arrow] (generate) -- (test);

\draw[arrow]
    (test)
    -- node[
        above left,
        font=\scriptsize\bfseries
    ] {FAIL}
    (diagnose);

\draw[arrow] (diagnose) -- (repair);
\draw[arrow] (repair) -- (retest);

\draw[arrow]
    (retest.west)
    -- ++(-7mm,0)
    |- (test.west);

\draw[arrow]
    (test)
    -- node[
        above right,
        font=\scriptsize\bfseries
    ] {PASS}
    (security);

\draw[arrow] (security) -- (proceed);

\end{tikzpicture}

\caption{
\textbf{Graph engineering:} explicit workflow structure governing
transitions, conditional branches, dependencies, recovery paths, and
progression conditions
~\protect\citep{openai2025practicalagents,anthropic2024effectiveagents}.
}
\label{fig:graph_engineering}
\end{figure}
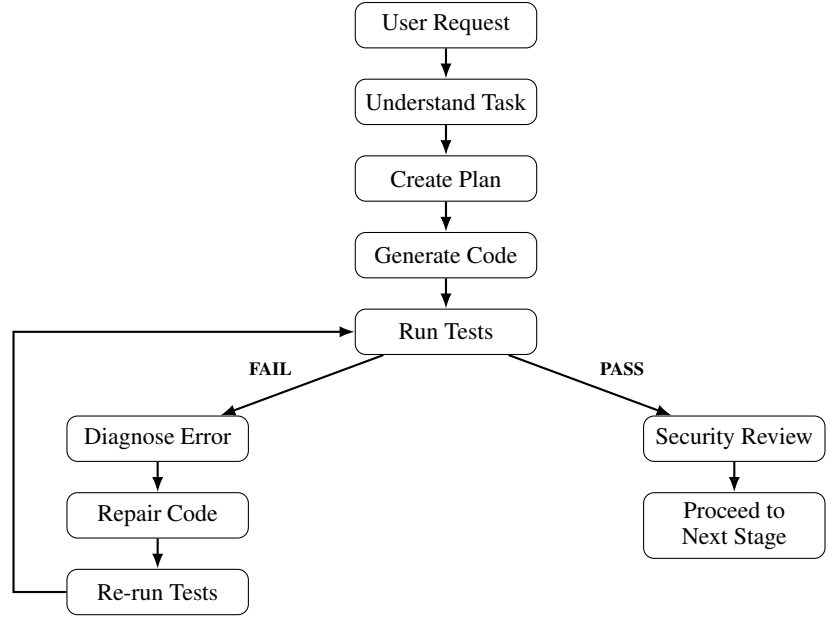

\paragraph{Graph Engineering.}
We use \emph{graph engineering} to denote the explicit organization of an agentic workflow into nodes, transitions, branches, dependencies, approvals, recovery paths, and terminal states. Workflow nodes may represent agents, actions, or execution states, while edges encode transitions, handoffs,
dependencies, and conditional control flow~\citep{openai2025practicalagents}. Structured agent workflows similarly support routing, parallel execution, and orchestrator--worker coordination~\citep{anthropic2024effectiveagents}. Graph engineering therefore determines admissible execution paths and the conditions governing progression, recovery, and termination.
Figure~\ref{fig:graph_engineering} illustrates these semantics: a passing test permits forward progression, whereas failure activates an explicit diagnosis--repair path before testing is repeated.

\begin{figure}[t]
\centering

\begin{tikzpicture}[
    node distance=4mm,
    every node/.style={
        font=\small,
        align=center
    },
    box/.style={
        draw,
        rounded corners,
        minimum width=22mm,
        minimum height=6mm
    },
    arrow/.style={
        -{Latex[length=2mm]},
        thick
    }
]

\node[box] (generate) {Generate};
\node[box, below=of generate] (execute) {Execute};
\node[box, below=of execute] (verify) {Verify};

\node[
    box,
    below left=8mm and 13mm of verify
] (bounds) {Recovery Bounds\\Available?};

\node[
    box,
    below=of bounds
] (diagnose) {Diagnose};

\node[
    box,
    below=of diagnose
] (repair) {Repair /\\Re-plan};

\node[
    box,
    below=of repair
] (reexecute) {Retry};

\node[
    box,
    below right=8mm and 13mm of verify
] (proceed) {Proceed};

\node[
    box,
    right=8mm of bounds
] (failure) {Auditable\\Failure};

\draw[arrow] (generate) -- (execute);
\draw[arrow] (execute) -- (verify);

\draw[arrow]
    (verify)
    -- node[
        above right,
        font=\scriptsize\bfseries
    ] {PASS}
    (proceed);

\draw[arrow]
    (verify)
    -- node[
        above left,
        font=\scriptsize\bfseries
    ] {FAIL}
    (bounds);

\draw[arrow]
    (bounds)
    -- node[
        left,
        font=\scriptsize\bfseries
    ] {YES}
    (diagnose);

\draw[arrow]
    (bounds)
    -- node[
        above,
        font=\scriptsize\bfseries
    ] {NO}
    (failure);

\draw[arrow] (diagnose) -- (repair);
\draw[arrow] (repair) -- (reexecute);

\draw[arrow]
    (reexecute.west)
    -- ++(-7mm,0)
    |- (verify.west);

\end{tikzpicture}

\caption{
\textbf{Loop engineering:} bounded diagnosis, repair or re-planning,
retry, and re-verification driven by execution or verification feedback
~\protect\citep{openai2026agentssdk,anthropic2024effectiveagents}.
}
\label{fig:loop_engineering}
\end{figure}
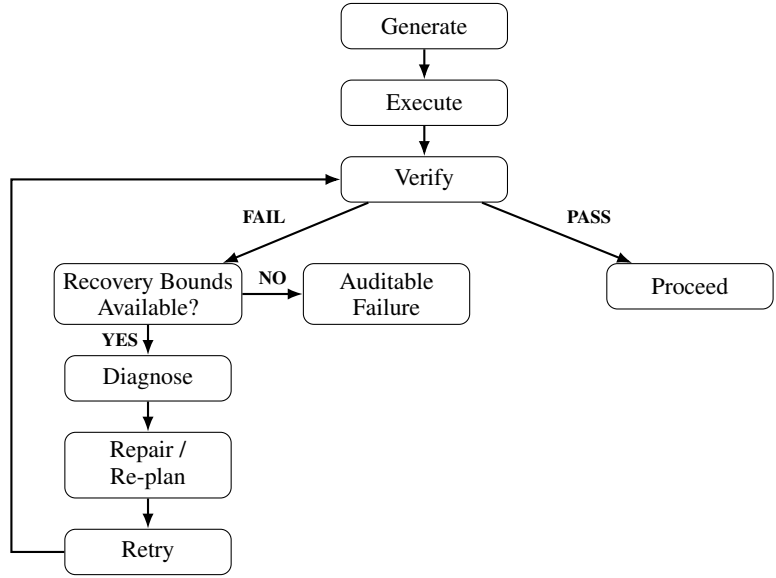

\paragraph{Loop Engineering.}
We use \emph{loop engineering} to denote the bounded iterative process through which an agent acts, observes execution or evaluation feedback, diagnoses failures, revises its actions or outputs, and retries. Agent runtimes commonly alternate between model reasoning and tool interaction
until completion, handoff, approval, or another stopping condition~\citep{openai2026agentssdk}. Evaluator--optimizer workflows similarly use evaluation feedback to guide iterative refinement under
explicit stopping conditions~\citep{anthropic2024effectiveagents}. Loop engineering specifies when recovery is triggered, how failure evidence informs diagnosis and repair or re-planning, whether another attempt is permitted, and when recovery must terminate. As shown in Fig.~\ref{fig:loop_engineering}, failed verification enters bounded diagnosis, repair or re-planning, retry, and re-verification while the applicable recovery bounds remain available; otherwise, execution terminates in an auditable failure state.

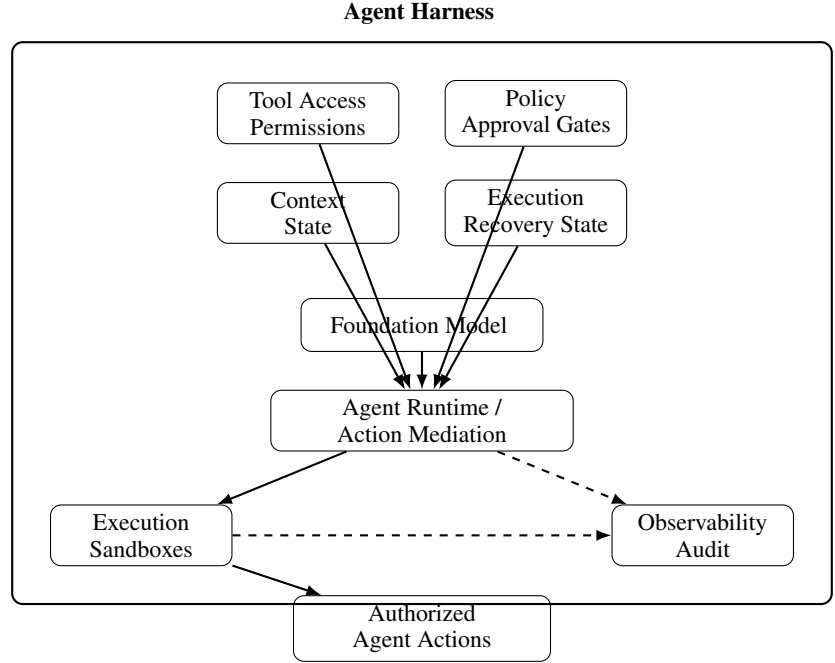
\begin{figure}[t]
\centering

\begin{tikzpicture}[
    node distance=5mm and 6mm,
    every node/.style={
        font=\small,
        align=center
    },
    box/.style={
        draw,
        rounded corners,
        minimum width=24mm,
        minimum height=7mm
    },
    widebox/.style={
        draw,
        rounded corners,
        minimum width=32mm,
        minimum height=7mm
    },
    outer/.style={
        draw,
        rounded corners,
        thick,
        inner sep=5mm
    },
    arrow/.style={
        -{Latex[length=2mm]},
        thick
    },
    obsarrow/.style={
        -{Latex[length=2mm]},
        thick,
        dashed
    }
]


\node[box] (tools) {
    Tool Access\\Permissions
};

\node[
    box,
    right=of tools
] (policy) {
    Policy\\Approval Gates
};

\node[
    box,
    below=of tools
] (state) {
    Context\\State
};

\node[
    box,
    right=of state
] (recovery) {
    Execution\\Recovery State
};


\coordinate (controlmid)
    at ($(state.south)!0.5!(recovery.south)$);

\node[
    widebox,
    below=7mm of controlmid
] (llm) {
    Foundation Model
};

\node[
    widebox,
    below=of llm,
    minimum width=40mm
] (runtime) {
    Agent Runtime /\\
    Action Mediation
};


\node[
    box,
    below left=7mm and 5mm of runtime
] (sandbox) {
    Execution\\Sandboxes
};

\node[
    box,
    below right=7mm and 5mm of runtime
] (observe) {
    Observability\\Audit
};

\node[
    widebox,
    below=10mm of runtime,
    yshift=-9mm,
    minimum width=34mm
] (actions) {
    Authorized\\Agent Actions
};


\node[
    outer,
    fit=(tools)(policy)(state)(recovery)(llm)(runtime)(sandbox)(observe)
] (harness) {};

\node[
    above=1mm of harness.north,
    font=\small\bfseries
] {
    Agent Harness
};


\draw[arrow] (tools) -- (runtime);
\draw[arrow] (policy) -- (runtime);
\draw[arrow] (state) -- (runtime);
\draw[arrow] (recovery) -- (runtime);

\draw[arrow] (llm) -- (runtime);

\draw[arrow] (runtime) -- (sandbox);
\draw[arrow] (sandbox) -- (actions);

\draw[obsarrow] (runtime) -- (observe);
\draw[obsarrow] (sandbox) -- (observe);

\end{tikzpicture}

\caption{
\textbf{Agent harness engineering:} controlled mediation of model-generated
actions through capability authorization, policy and approval gates, state,
sandboxed execution, recovery state, observability, and audit
~\protect\citep{openai2026sandboxagents,anthropic2025effectiveharnesses}.
}
\label{fig:agent_harness}
\end{figure}

\paragraph{Agent Harness Engineering.}
We use \emph{agent harness engineering} to denote the controlled runtime
and policy boundary that mediates model invocation, capability access, external actions, state, and execution environments. The harness governs tool routing, permissions, approvals, execution state, tracing, and sandboxed execution~\citep{openai2026sandboxagents}. General-purpose agent
harnesses similarly provide mechanisms through which models obtain context, invoke tools, execute actions, and maintain state across extended tasks~\citep{anthropic2025effectiveharnesses}. Agent harness engineering therefore determines which capabilities may be invoked, which resources are
accessible, where actions execute, what state is retained, and how execution is authorized, isolated, observed, and audited. Figure~\ref{fig:agent_harness} summarizes these control surfaces and shows how the harness mediates model-generated actions before external execution.

\subsection{Unified Graph--Loop--Harness Architecture}
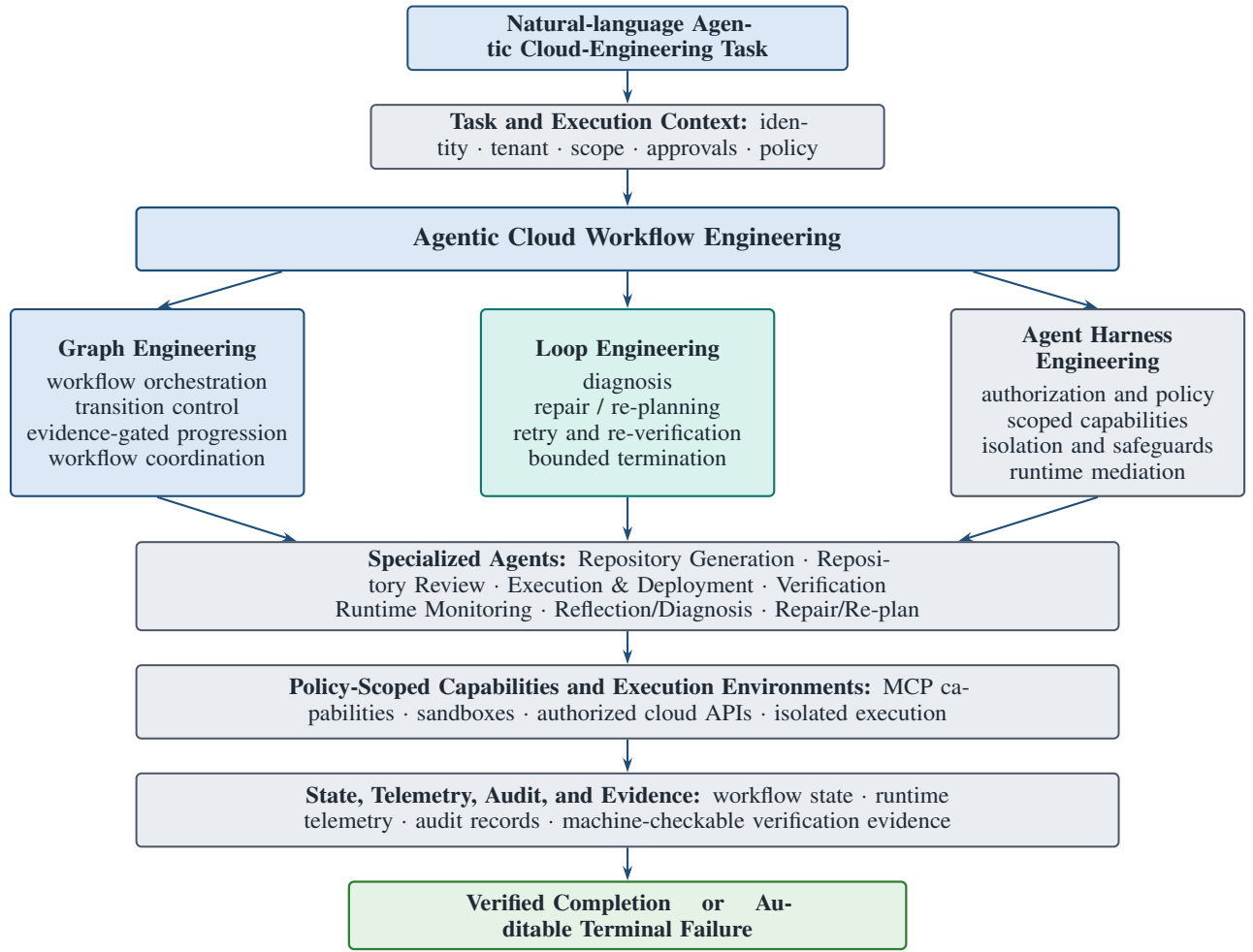
\begin{figure*}[t]
\centering
\hspace*{-2mm}%

\begin{tikzpicture}[
    node distance=4.5mm,
    every node/.style={
        align=center,
        font=\small,
        text=ACEInk
    },
    input/.style={
        draw=ACEGraph,
        fill=ACEGraphFill,
        rounded corners=2pt,
        line width=0.75pt,
        minimum width=6.0cm,
        text width=5.6cm,
        minimum height=0.82cm,
        font=\small\bfseries
    },
    context/.style={
        draw=ACEControl,
        fill=ACEControlFill,
        rounded corners=2pt,
        line width=0.75pt,
        minimum width=7.0cm,
        text width=6.6cm,
        minimum height=0.86cm,
        font=\small
    },
    pillar/.style={
        rounded corners=2pt,
        line width=0.8pt,
        minimum width=4.0cm,
        text width=3.6cm,
        minimum height=2.55cm,
        font=\small
    },
    layer/.style={
        draw=ACEControl,
        fill=ACEControlFill,
        rounded corners=2pt,
        line width=0.75pt,
        minimum width=13.4cm,
        text width=13.0cm,
        minimum height=1.00cm,
        font=\small
    },
    outcome/.style={
        draw=ACEVerify,
        fill=ACEVerifyFill,
        rounded corners=2pt,
        line width=0.85pt,
        minimum width=7.6cm,
        text width=7.2cm,
        minimum height=0.92cm,
        font=\small\bfseries
    },
    mainarrow/.style={
        -{Stealth[length=2mm,width=1.3mm]},
        draw=ACEGraph,
        line width=0.8pt
    }
]


\node[input] (task) {
    Natural-language Agentic Cloud-Engineering Task
};


\node[
    context,
    below=of task
] (context) {
    \textbf{Task and Execution Context:}
    identity $\cdot$ tenant $\cdot$ scope $\cdot$
    approvals $\cdot$ policy
};


\node[
    draw=ACEGraph,
    fill=ACEGraphFill,
    rounded corners=2pt,
    line width=0.85pt,
    below=5mm of context,
    minimum width=13.4cm,
    text width=13.0cm,
    minimum height=0.86cm,
    font=\normalsize\bfseries
] (framework) {
    Agentic Cloud Workflow Engineering
};


\node[
    pillar,
    draw=ACEGraph,
    fill=ACEGraphFill,
    below left=5mm and 4.4cm of framework.south
] (graph) {
    \textbf{Graph Engineering}\\[1mm]
    workflow orchestration\\
    transition control\\
    evidence-gated progression\\
    workflow coordination
};


\node[
    pillar,
    draw=ACELoop,
    fill=ACELoopFill,
    below=5mm of framework.south
] (loop) {
    \textbf{Loop Engineering}\\[1mm]
    diagnosis\\
    repair / re-planning\\
    retry and re-verification\\
    bounded termination
};


\node[
    pillar,
    draw=ACEControl,
    fill=ACEControlFill,
    below right=5mm and 4.4cm of framework.south
] (harness) {
    \textbf{Agent Harness Engineering}\\[1mm]
    authorization and policy\\
    scoped capabilities\\
    isolation and safeguards\\
    runtime mediation
};


\node[
    layer,
    below=6mm of loop
] (agents) {
    \textbf{Specialized Agents:}
    Repository Generation $\cdot$ Repository Review $\cdot$
    Execution \& Deployment $\cdot$ Verification\\[-0.2mm]
    Runtime Monitoring $\cdot$ Reflection/Diagnosis $\cdot$
    Repair/Re-plan
};


\node[
    layer,
    below=of agents
] (execution) {
    \textbf{Policy-Scoped Capabilities and Execution Environments:}
    MCP capabilities $\cdot$ sandboxes $\cdot$
    authorized cloud APIs $\cdot$ isolated execution
};


\node[
    layer,
    below=of execution
] (evidence) {
    \textbf{State, Telemetry, Audit, and Evidence:}
    workflow state $\cdot$ runtime telemetry $\cdot$
    audit records $\cdot$ machine-checkable verification evidence
};


\node[
    outcome,
    below=of evidence
] (outcome) {
    Verified Completion
    \quad or \quad
    Auditable Terminal Failure
};


\draw[mainarrow]
    (task) -- (context);

\draw[mainarrow]
    (context) -- (framework);

\draw[mainarrow]
    ($(framework.south west)+(2.0cm,0)$)
    -- (graph.north);

\draw[mainarrow]
    (framework.south)
    -- (loop.north);

\draw[mainarrow]
    ($(framework.south east)+(-2.0cm,0)$)
    -- (harness.north);

\draw[mainarrow]
    (graph.south)
    -- ($(agents.north west)+(2.2cm,0)$);

\draw[mainarrow]
    (loop.south)
    -- (agents.north);

\draw[mainarrow]
    (harness.south)
    -- ($(agents.north east)+(-2.2cm,0)$);

\draw[mainarrow]
    (agents) -- (execution);

\draw[mainarrow]
    (execution) -- (evidence);

\draw[mainarrow]
    (evidence) -- (outcome);

\end{tikzpicture}

\caption{High-level organization of \textbf{Agentic Cloud Workflow Engineering}, integrating graph-controlled workflow progression, bounded recovery, policy-constrained capability use and execution, and machine-checkable evidence for verified completion or bounded terminal failure.
}
\label{fig:cloud-framework}
\vspace{-3mm}
\end{figure*}

The three concerns compose a single execution framework rather than sequential lifecycle stages. Figure~\ref{fig:cloud-framework} shows this composition within \textbf{Agentic Cloud Workflow Engineering}. A natural-language task is first bound to identity, tenant, scope, approvals,
and policy. Graph engineering governs admissible workflow progression, loop engineering governs bounded recovery, and agent harness engineering governs authorized and isolated capability use. Their composition governs agent coordination, recovery, and interaction with policy-scoped capabilities and execution environments. Workflow state, runtime telemetry, audit records,
and machine-checkable verification evidence provide the control and verification substrate for progression and termination. A workflow therefore terminates either in verified completion or in an auditable failure state when the required verification conditions remain unsatisfied after the
applicable recovery bounds are exhausted.

\end{document}